\documentclass[aps, prb, superscriptaddress, twocolumn]{revtex4-1}
\usepackage[english]{babel}
\usepackage{amsmath,amsthm}
\usepackage{amsfonts}
\usepackage[pdfborder={0 0 0}, colorlinks=true, urlcolor=blue, linkcolor=blue, citecolor=blue]{hyperref}
\usepackage{color}
\usepackage{graphicx}
\usepackage{float}
\usepackage{subfigure}

\theoremstyle{definition}

\theoremstyle{remark}

\begin{document}

\title{Emergence and stability of spontaneous vortex lattices in exciton-polariton condensates}
\author{F. X. Sun}
\affiliation{State Key Laboratory for Mesoscopic Physics  and  Collaborative Innovation Center of Quantum Matter, School of Physics,  Peking University, Beijing 100871, China}
\affiliation{Collaborative Innovation Center of Extreme Optics, Shanxi University, Taiyuan, Shanxi 030006, China}
\author{Z. X. Niu}
\affiliation{Department of Physics, Renmin University of China, Beijing 100872, China}
\author{Q. H. Gong}
\affiliation{State Key Laboratory for Mesoscopic Physics  and  Collaborative Innovation Center of Quantum Matter, School of Physics,  Peking University, Beijing 100871, China}
\affiliation{Collaborative Innovation Center of Extreme Optics, Shanxi University, Taiyuan, Shanxi 030006, China}
\affiliation{Beijing Academy of Quantum Information Sciences, Haidian District, Beijing 100193, China}
\affiliation{Nano-optoelectronics Frontier Center of the Ministry of Education, Beijing 100871, China}
\author{Q. Y. He}
\email{qiongyihe@pku.edu.cn}
\affiliation{State Key Laboratory for Mesoscopic Physics  and  Collaborative Innovation Center of Quantum Matter, School of Physics,  Peking University, Beijing 100871, China}
\affiliation{Collaborative Innovation Center of Extreme Optics, Shanxi University, Taiyuan, Shanxi 030006, China}
\affiliation{Beijing Academy of Quantum Information Sciences, Haidian District, Beijing 100193, China}
\affiliation{Nano-optoelectronics Frontier Center of the Ministry of Education, Beijing 100871, China}
\author{W. Zhang}
\email{wzhangl@ruc.edu.cn}
\affiliation{Department of Physics, Renmin University of China, Beijing 100872, China}
\affiliation{Beijing Key Laboratory of Opto-Electronic Functional Materials and Micro-Nano Devices, Renmin University of China, Beijing 100872, China}

\begin{abstract}
	The spontaneous formation of lattice structure of quantized vortices is a characteristic feature of superfluidity in closed systems under thermal equilibrium. In exciton-polariton Bose-Einstein condensate, which is a typical example of macroscopic quantum state in open systems, spontaneous vortex lattices have also been proposed by not yet observed. Here, we take into account the finite decay rate of exciton reservoir, and theoretically investigate the vortex structures in circularly pumped polariton Bose-Einstein condensate. Our results show that a decreasing reservoir decay rate can reduce the number of vortices and destabilize the lattice structure, hence is unfavorable to the formation and observation of vortex lattices. These detrimental effects can be prevailed by applying an external angular momentum.
\end{abstract}

\maketitle

\section{Introduction}
As one of the most fascinating quantum phenomena, superfluidity remained to be the frontier of modern physics and attracted constant and improving attention since its discovery in 1930s.~\cite{KAPITZA1938, ALLEN1938} With the ability of carrying current without any dissipation, such an exotic state of matter not only manifests the role of quantum coherence, but also reveals fascinating directions of potential applications. For Bose systems, superfluidity has been observed and studied in liquid $^4$He~\cite{KAPITZA1938, ALLEN1938}, ultracold quantum gases of bosonic atoms,~\cite{ketterle-00} and magnons in magnetic compounds~\cite{tanaka-00}. In all these examples, the bosonic particles contribute a macroscopic occupation of the single-particle ground state to form a Bose-Einstein condensate (BEC), where the interaction plays a key role in the presence of superfluidity.~\cite{bogoliubov-47} 

The realization of exciton-polariton BEC, or equivalently polariton BEC for short, adds another member in the family of BEC and provides us the opportunity to study such a macroscopic quantum state in open systems.~\cite{Deng199,Kasprzak2006,Balili1007,kim2008gaas} The exciton is a quasiparticle consisting of an electron and a hole, which can be confined in a two-dimensional (2D) geometry in quantum wells embedded in optical microcavity. When the electron-photon coupling is strong, the exciton and the cavity photon modes are both dressed to form new eigenstates of this hybrid system. The new bosonic quasiparticles, referred as polariton, can in principle condense into the single-particle ground state. Thanks to the extremely light effective mass of polaritons, the polariton BEC can have a transition temperature $T_c$ as high as room temperature.~\cite{Christopoulos-07, Baumberg-08}

Another distinctive feature of polariton BEC is that it is indeed an open system under external pumping and decay. One of the decay channel is the leakage of photons from the cavity. Besides, the exciton component of polariton can also decay via radiative and nonradiative processes. As a consequence, polariton BEC is truly a dynamical steady state instead of a thermal equilibrium state. A natural question then arises: can the polariton BEC also support superfluidity? On one hand, the Bogoliubov excitation spectrum is believed to deviate from the linear dispersion as in the conventional BEC,~\cite{wouters2007excitations} which compromises the Landau criterion for the critical velocity of superfluid. Further theoretical studies suggest that the superfluid order can only exist in such a open-driven system under strong anisotropic confinement.~\cite{altman-15, altman-16} On the other hand, experimental results of the suppression of scattering from impurities~\cite{Carusotto-04, Amo2009, Amo2009-2}, and the long lifetime of induced vortices and imprinted vortex configuration suggest the existence of superfluidity.~\cite{Lagoudakis2008, Sanvitto2010, roumpos2011single, Bramatireview, Bramati2015}

One characteristic feature of superfluid is the spontaneous emergence of lattice structure of quantized vortices when subjected to finite angular momentum. After the first proposal in the context of type-II superconductor where electron pairs can move without dissipation to form charged superfluid,~\cite{Abrikosov-57} vortex lattice has been experimentally observed and considered to be a strong evidence for superfluidity in Bose and Fermi atomic condensates.~\cite{Zwierlein2005, Abo-Shaeer476} For polariton BEC, theoretical proposals have been made to generate vortex lattices or chains by engineering the pumping potential.~\cite{Liew-08, keeling2008spontaneous, Borgh2010, Gorbach-10, Borgh2012, padhi2015vortex, chen2017quantum} Among these works, Keeling and Berloff employ the adiabatic approximation by assuming the decay rate $\gamma_R$ of the exciton reservoir is much larger than that of the polariton $\gamma_c$, and propose to stabilize a spontaneously formed vortex lattice with a finite sized circular pumping laser.~\cite{keeling2008spontaneous} Within this framework, the reservoir can be adiabatically eliminated and the system is described by a generalized Gross-Pitaevskii formalism with complexed decay and gain terms. Under the same approximation, subsequent works generalize the configuration to disordered environment,~\cite{Borgh2012} noncircular geometry,~\cite{Borgh2012} spinor polariton condensates,~\cite{Borgh2010} and rotating systems.~\cite{padhi2015vortex,chen2017quantum}

In this work, we go beyond the adiabatic approximation and investigate the effect of reservoir decay on the formation of vortex lattices in a circularly pumped polariton BEC. We find that by reducing the reservoir decay from the limiting condition $\gamma_R \gg \gamma_c$, fewer vortices can be generated in the system, while the lattice structure can still be stabilized until $\gamma_R \gtrsim \gamma_c$. If the reservoir decay is further decreased, the vortex lattice starts to melt and eventually liquify when $\gamma_R \ll \gamma_c$. In this limiting regime, the vortex lattice can be reestablished by applying an overall external angular momentum. Our results suggest that a large reservoir decay and an external angular momentum favor the generation and observation of vortex lattices in polariton BEC. 

The remainder of this manuscript is organized as follows. In Sec.~\ref{sec:model}, we present the mean-field formalism to incorporate the reservoir decay and the simplification under the adiabatic approximation. The results of vortex lattices are discussed in Sec.~\ref{sec:result}, where numerical simulations with different reservoir decays are compared. Finally, we summarize in Sec.~\ref{sec:summary}.

%%%%%%%%%
\section{Models}
\label{sec:model}

To investigate the effect of finite decay rate of exciton reservoir, we employ a mean-field treatment for the polariton BEC and adopt the open-dissipative Gross-Pitaevskii equation (ODGPE). This formalism is introduced in Ref.~\cite{wouters2007excitations}, and commonly used in subsequent works to successfully explain a large number of experiments in exciton-polariton condensates.~\cite{Lagoudakis2008, roumpos2011single, manni2013spontaneous, sanvitto2016road, berloff2017realizing} Within this framework, the mean-field wave function $\psi$ of the polariton condensate and the density of reservoir $n_R$ satisfy a coupled equation set
\begin{eqnarray}
	i\frac{\partial\psi}{\partial t}&=&[ -\frac{\hbar\nabla^2}{2m}+V(r)+\frac{g_c}{\hbar}|\psi|^2+\frac{g_R}{\hbar}n_R+\frac{i}{2}(Rn_R-\gamma_c) ]\psi, \nonumber\\
	\frac{\partial n_R}{\partial t}&=&P-(\gamma_R+R|\psi|^2)n_R.
	\label{eqn:ODGPE}
\end{eqnarray}
Here, $P$ is the exciton creation rate determined by the external pumping, $R$ is the rate of stimulated scattering from the reservoir to the condensate, $m$ is the effective mass of polariton, and $V(r)$ is an external potential. Both polariton and reservoir are lossy with decay rates $\gamma_c$ and $\gamma_R$, respectively, and are repulsively interacted via polariton-polariton repulsion $g_c$ and polariton-reservoir interaction $g_R$. Previous analysis show that a stable polariton condensate exists with the condensate density $|\psi_{ss}|^2=P/\gamma_c-\gamma_R/R$ and the reservoir density $n_R^{ss}=\gamma_c/R$ when the uniform pumping $P>P_{th} \equiv \gamma_c\gamma_R/R$.~\cite{wouters2007excitations,xue2014creation,kulczykowski2017phase}

It is usually desirable to derive a dimensionless form of the coupled equations~(\ref{eqn:ODGPE}), from which some universal features can be revealed. Here, we consider an external harmonic trapping potential $V(r)=(1/2)m\omega^2r^2$ with $\omega$ the oscillator frequency, and rescale the equations~(\ref{eqn:ODGPE}) using the length unit $\ell=\sqrt{\hbar/m\omega}$ and time unit $t_0=2/\omega$. Then we get
\begin{eqnarray}
	i\frac{\partial\psi}{\partial t}&=&\left[ -\nabla^2+r^2+g'_c|\psi|^2+g'_Rn_R+\frac{i}{2}(R'n_R-\gamma'_c) \right]\psi, \nonumber\\
	\frac{\partial n_R}{\partial t}&=&P'-(\gamma'_R+R'|\psi|^2)n_R.
	\label{eqn:ODGPE_rescaled}
\end{eqnarray}
Notice that in the expressions above, the wave function $\psi$, coordinate $r$, time $t$ and density $n_R$ are all replaced by their dimensionless counterparts, while other parameters are defined as $P'=2P \ell^2/\omega$, $R'=2R/\omega \ell^2$, $\gamma'_c=2\gamma_c/\omega$, $\gamma'_R=2\gamma_R/\omega$, $g'_c=2g_c/\hbar\omega \ell^2$, and $g'_R=2g_R/\hbar\omega \ell^2$. In the following discussion, we focus on this dimensionless form and perform numerical simulations to obtain time evolution of the polariton BEC.

If the parameters in Eq.~(\ref{eqn:ODGPE_rescaled}) satisfy the conditions $\gamma'_R\gg \gamma'_c$ and $\gamma'_R\gg P'R'/(2\gamma'_c)$, we can adopt a so-called adiabatic approximation to eliminate the reservoir adiabatically and derive a decoupled equation for the polariton wave function~\cite{keeling2008spontaneous, Borgh2012, chen2013collective, moxley2016sagnac},
\begin{equation}
	i\frac{\partial\psi}{\partial t}=\left[ -\nabla^2+r^2+g|\psi|^2+i(\alpha-\sigma|\psi|^2) \right]\psi,
	\label{cGPE}
\end{equation}
where $g=g'_c-g'_RP'R'/{\gamma'}_R^2$ is the effective rate of repulsive polariton-polariton interaction, 
$\alpha=P'R'/(2\gamma'_R)-\gamma'_c/2$ is the effective pumping rate, and $\sigma=P'R'^2/(2{\gamma'}_R^2) $ is the effective rate of saturation loss. This so-called generalized Gross-Pitaevskii equation (GGPE) can be easily derived by setting $\partial n_R/\partial t=0$ and then plugging the result of $n_R$ into the first row of Eq.~(\ref{eqn:ODGPE_rescaled}). Note that there are also some sophisticated models which treat the reservoir or polariton scattering in a more delicate way by including the energy relaxation of polaritons or the stochastic noises from the interaction with environment~\cite{wouters2010energy, solnyshkov2014hybrid, wouters2009stochastic, chiocchetta2013non, carusotto2013quantum}. Most of them are just generalized from the ODGPE considered here, and the effects of adiabatic approximation can be analyzed analogously. 

%%%%%%%%%%
\section{Emergence and stability of vortex lattices}
\label{sec:result}

Under the framework of GGPE, previous studies predict that vortex lattices can be spontaneous emerged in polariton BEC when pumped by a circular spot $\alpha(r)=\alpha\Theta(R_P-r)$, where $\Theta$ is the unit step function and $R_P$ the cutoff radius.~\cite{keeling2008spontaneous} In this section, we study the effects of adiabatic approximation on the spontaneous vortex lattice by comparing solutions of ODGPE and GGPE under various circumstances. There are some earlier works analyzing the validity of adiabatic approximation on the exciton-polariton condensates~\cite{bobrovska2015adiabatic}, while the discussion on vortices and vortex lattices are still lack. 

\subsection{Spontaneous vortex lattices with the adiabatic approximation}\label{Sec:adiabatic}

First, we study the properties of vortex lattices in the exciton-polariton BEC under adiabatic approximation, where the simplified GGPE are equivalent to the ODGPE. According to Ref.~\cite{keeling2008spontaneous}, the spontaneous vortex lattices can be formed when the radius of the circular pumping laser exceeds the Thomas-Fermi radius of condensate $R_P>R_{TF}=\sqrt{3g\alpha/2\sigma}$. Here, we numerically solve the GGPE (\ref{cGPE}) to extract the time evolution of polariton BEC starting from a fixed initial state in the form of the Thomas-Fermi distribution
\begin{equation}
	\psi_0=\left\{
		\begin{array}{cc}
			\sqrt{\frac{3\alpha g-2\sigma r^2}{2\sigma g}}, & r<\sqrt{\frac{3g\alpha}{2\sigma}},\\
			0, & r\ge\sqrt{\frac{3g\alpha}{2\sigma}},
		\end{array}\right.
	\label{initial}
\end{equation}
and obtain similar results as in Ref.~\cite{keeling2008spontaneous}.

\begin{figure}[tbp]
	\centering
	\includegraphics[width=0.22\textwidth]{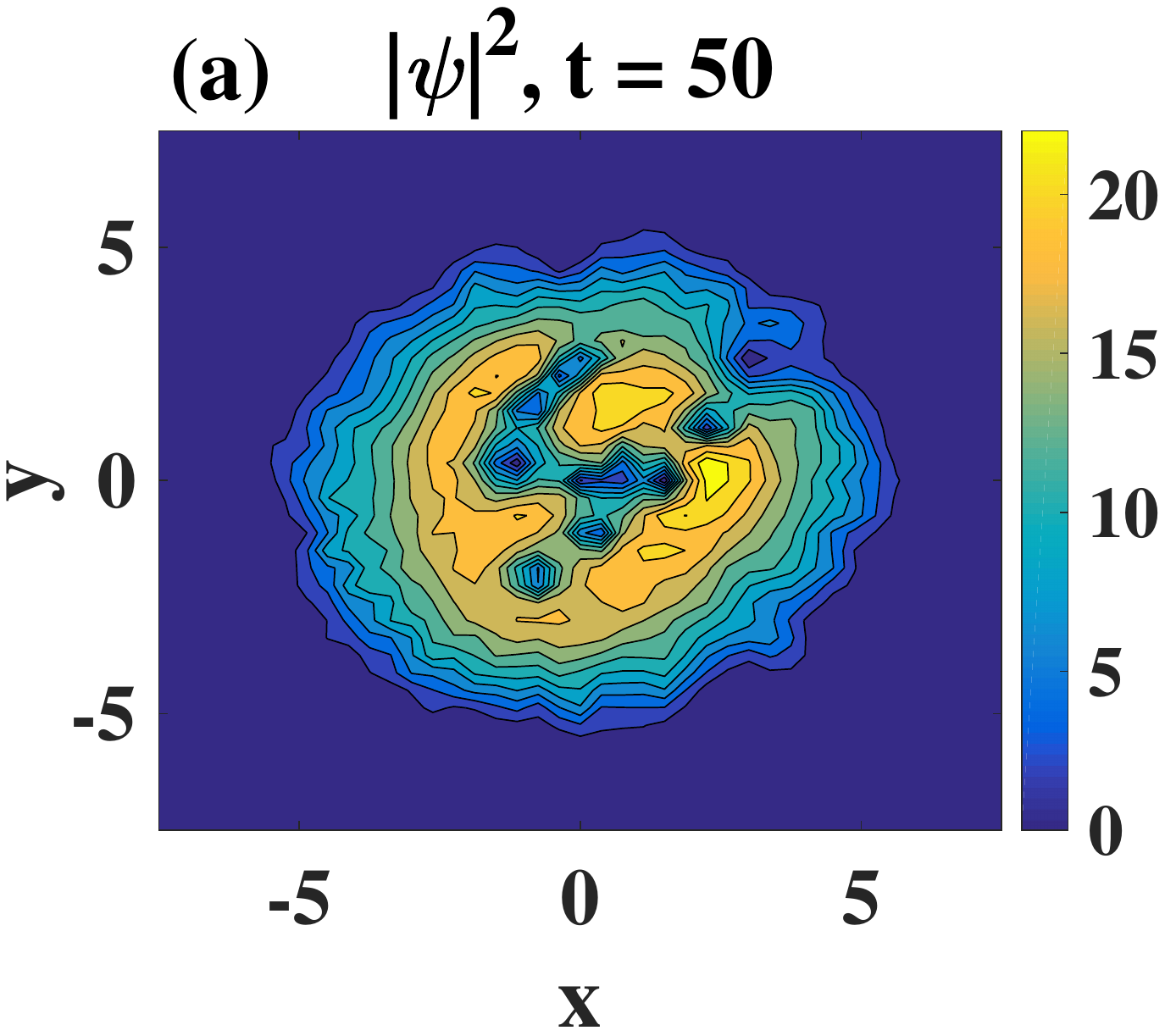}
	\includegraphics[width=0.22\textwidth]{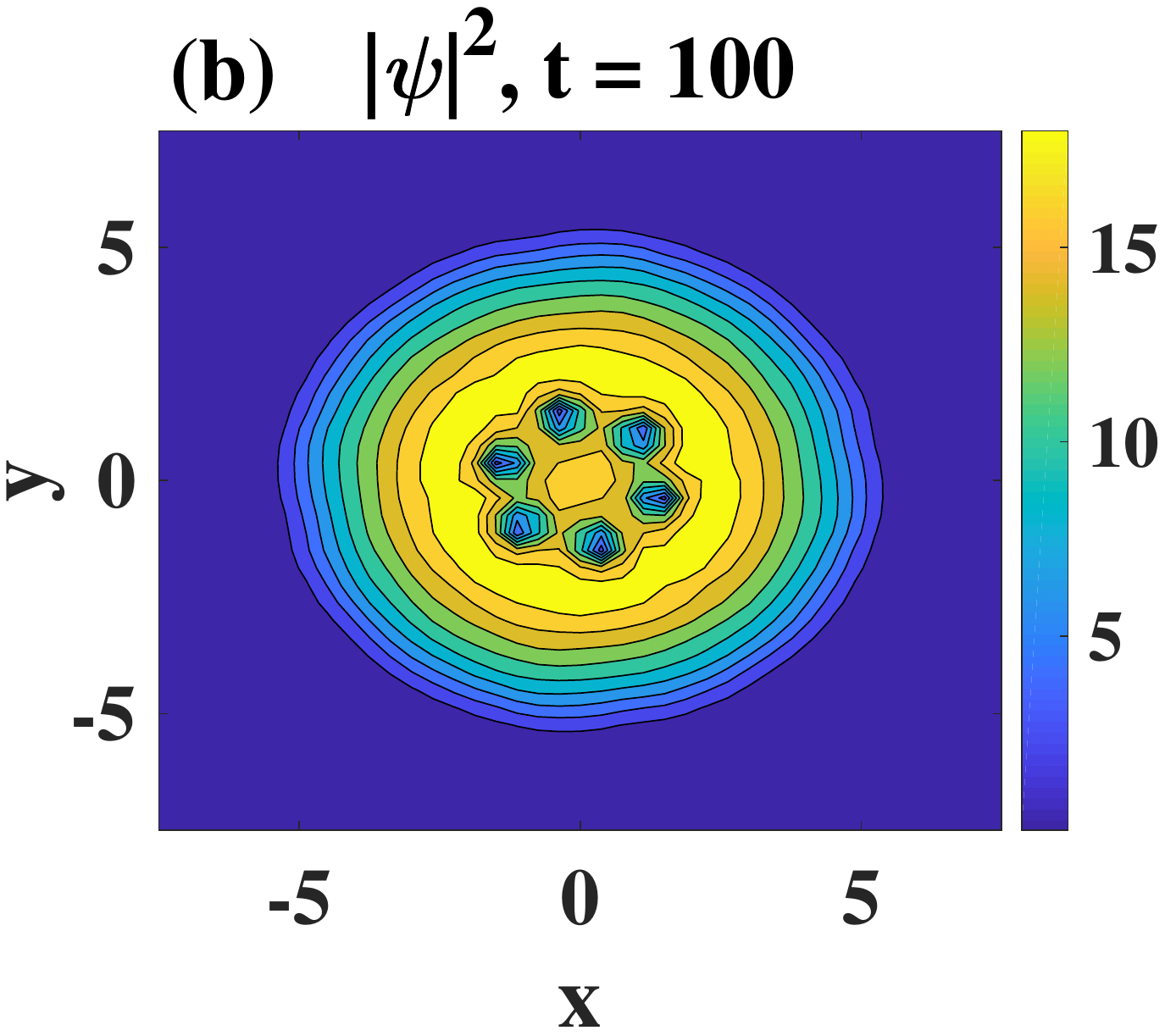}
	\\
	\vskip2mm
	\includegraphics[width=0.22\textwidth]{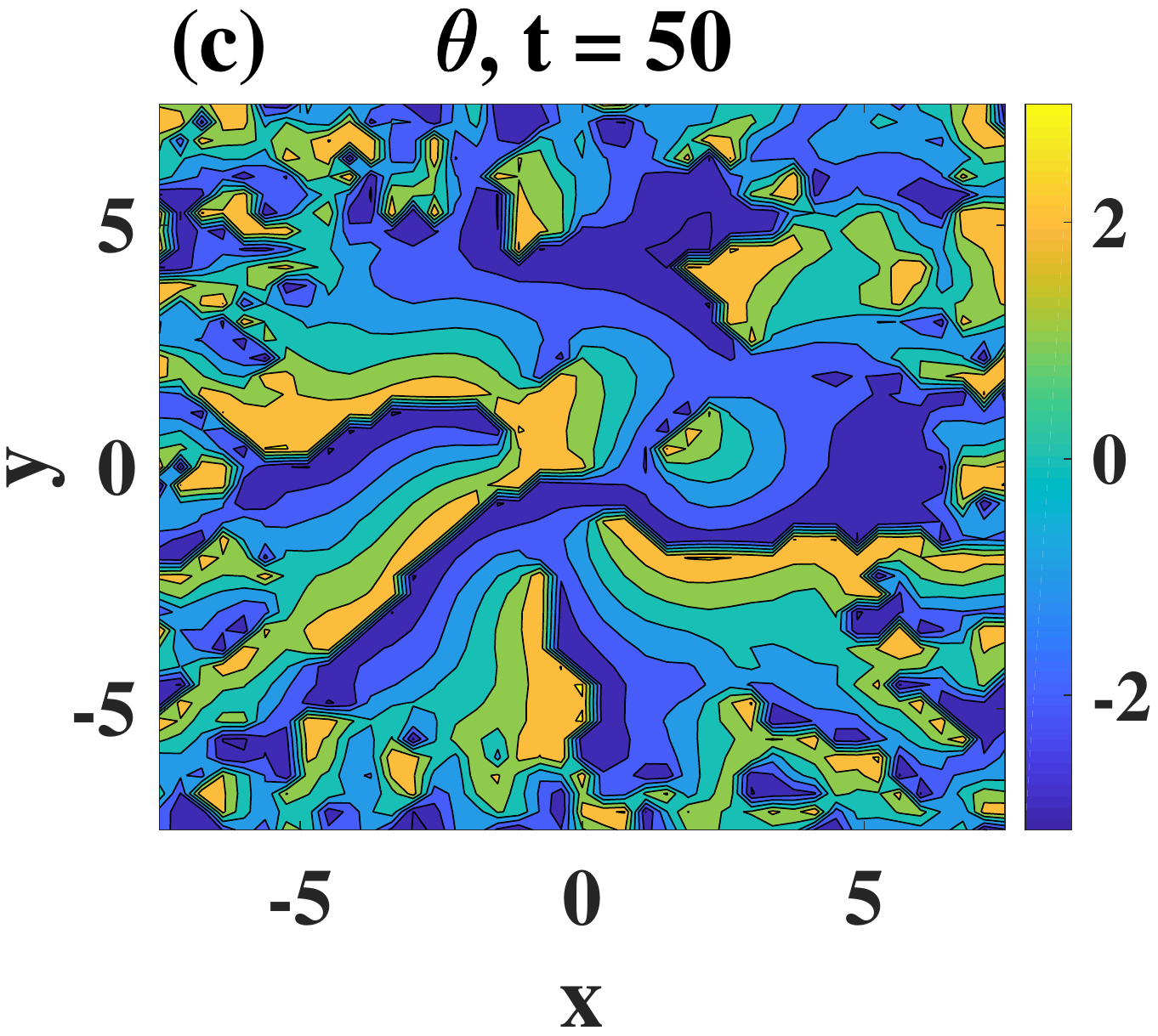}
	\includegraphics[width=0.22\textwidth]{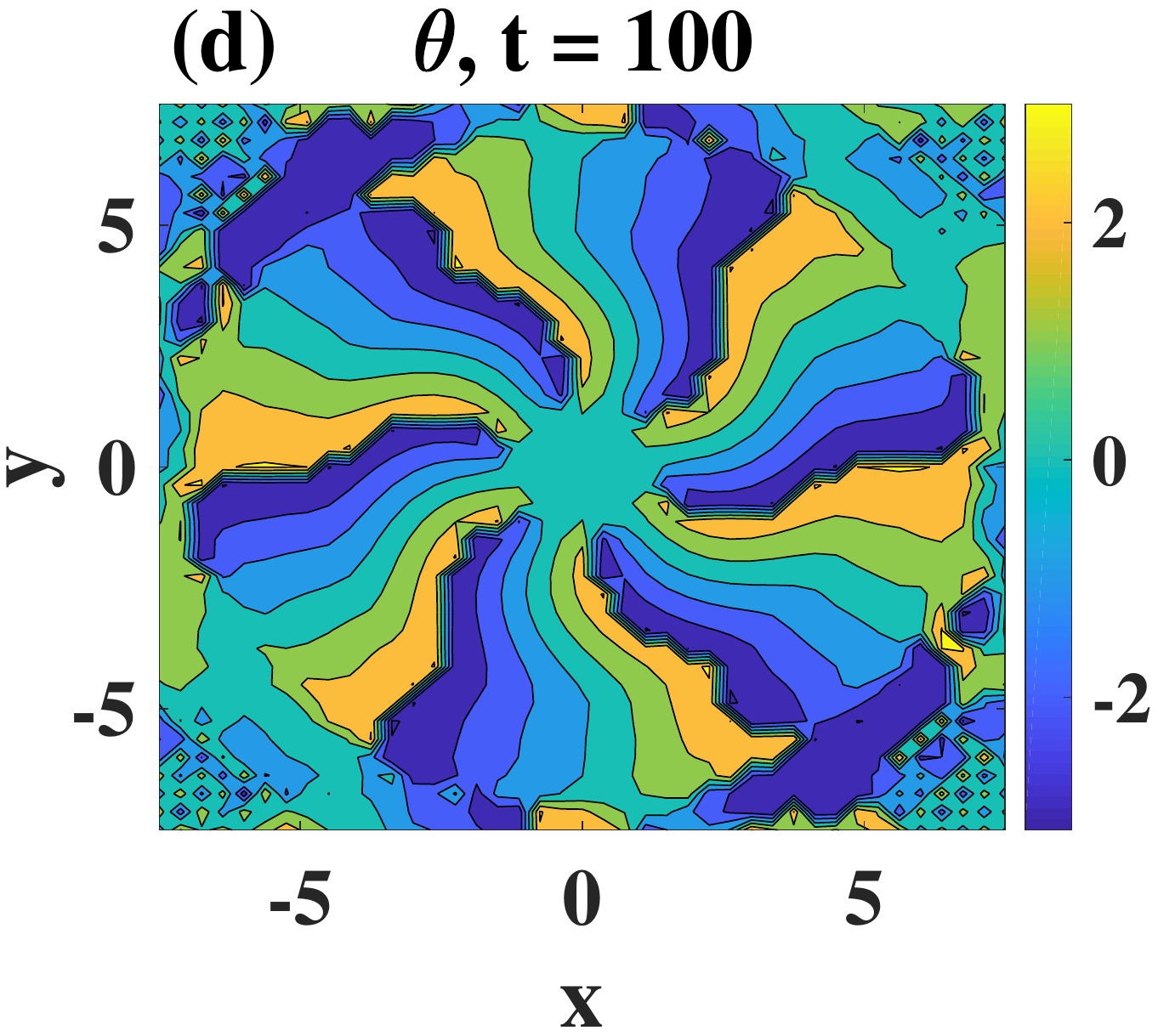}
	\\
	\vskip2mm
	\includegraphics[width=0.22\textwidth]{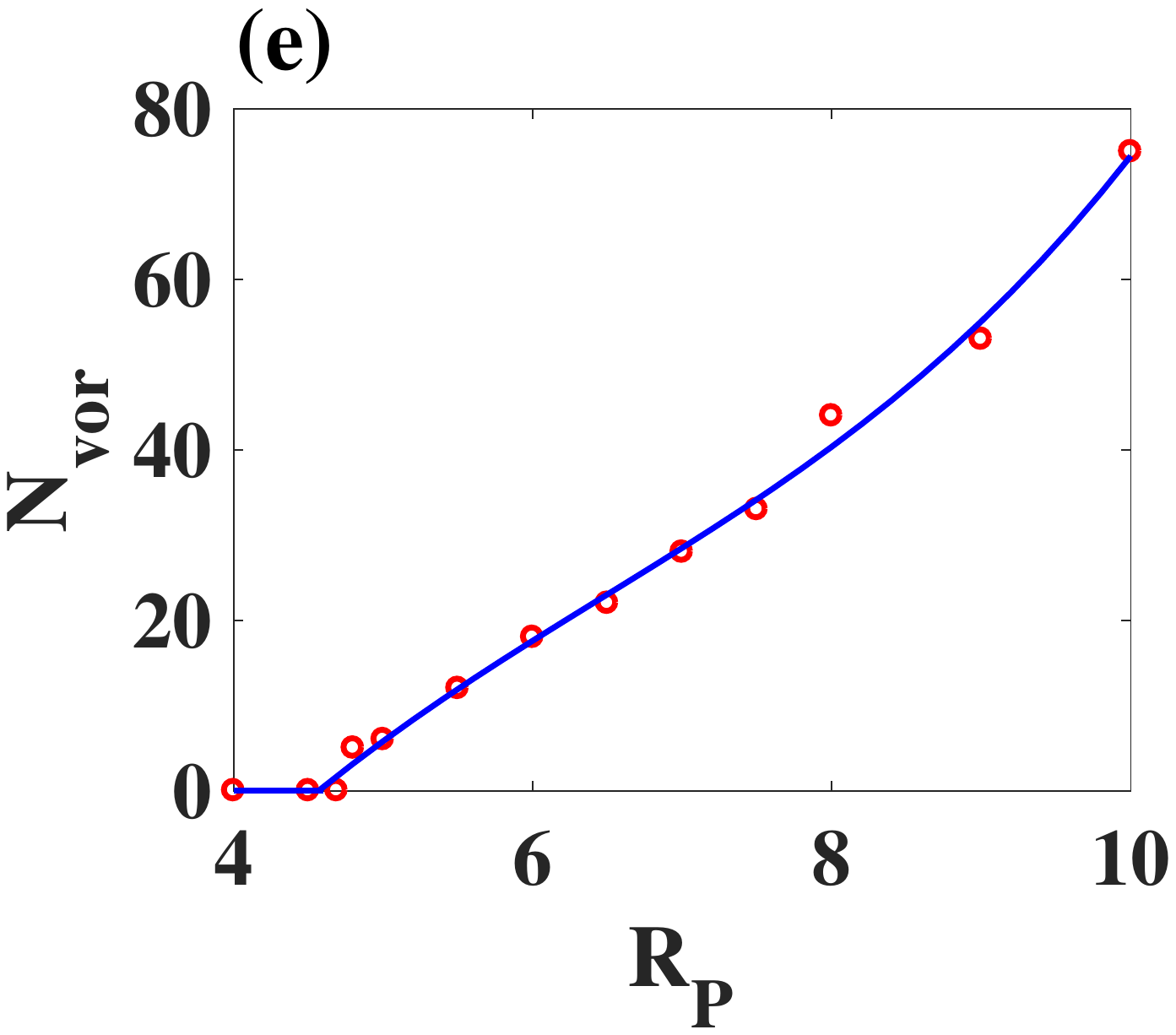}
	\includegraphics[width=0.22\textwidth]{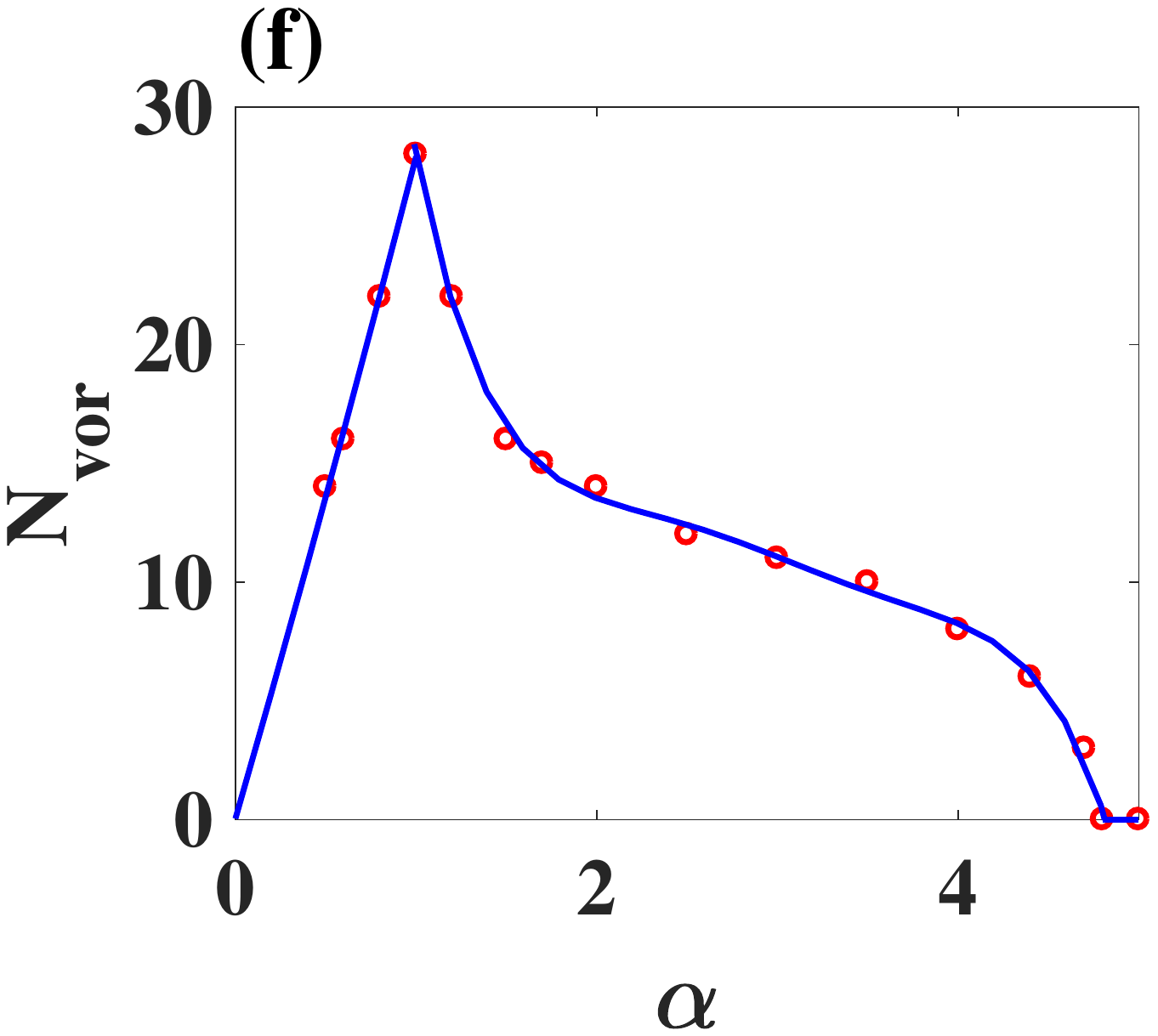}
	\caption{(Color online) Time evolution and steady-state solution of a polariton BEC obtained by numerically solving Eq.~(\ref{cGPE}) with initial condition Eq.~(\ref{initial}) for $\sigma=0.3$, $\alpha=4.4$, $g=1$ when the radius of the finite pumping spot is $R_P=5$. The panels (a) and (b) show the time evolution of condensate density, while (c) and (d) give the corresponding distributions of phase. Panels (e) and (f) show the number of vortices $N_{\rm vor}$ in the steady state by varying the laser radius $R_P$ and pumping rate $\alpha$, respectively, where the red circles are the numerical results obtained in the long-time limit and the blue solid lines are guide for the eye.}
	\label{fig:case6-GP}
\end{figure}

In Fig.~\ref{fig:case6-GP}, we show the emergence of vortices and vortex lattices via the distributions of number density [Figs.~\ref{fig:case6-GP}~(a) and \ref{fig:case6-GP}(b)] and phase [Figs.~\ref{fig:case6-GP}~(c) and \ref{fig:case6-GP} (d)] of the polariton condensates. The number of vortices with different pumping radius $R_P$ or pumping rate $\alpha$ has also been estimated in Fig.~\ref{fig:case6-GP}~(e) and (f), respectively. Notice that the number of vortices increases with the radius of the pumping laser, which is consistent with the discussion of Ref.~\cite{keeling2008spontaneous}. However, we do not observe a quadratic relation $N_{vor}\propto R_P^2$ predicted for large $R_P$,~\cite{keeling2008spontaneous} as the laser radius in our simulation is not large enough. In fact, we find that the vortex lattice becomes irregular and unstable for large $R_P$, as suggested in Ref.~\cite{keeling2008spontaneous}.

The number of vortices depends on the pumping rate $\alpha$ in a non-monotonic manner. When the pumping rate is elevated from zero, the number of vortices increase at first and then decrease to zero. This behavior indicates that the pumping rate $\alpha$ can affect the generation of vortices in two competing aspects. As shown in Eq.~(\ref{initial}), the Thomas-Fermi radius of the polariton BEC is related to the pumping rate as $R_{TF}\propto\sqrt{\alpha}$. In the region of small enough $\alpha$, the laser radius $R_P \gg R_{TF}$. The increase of pumping rate would favor the formation of polariton condensate and enhance the cloud size, such that more vortices can be accommodated, while the condition $R_P \gg R_{TF}$ remains valid. When the pumping rate is further elevated, the condensate Thomas-Fermi radius eventually becomes comparable to or even larger than the laser spot. The condition $R_P>R_{TF}$ is then compromised and the vortex formation is less favorable. 

\begin{figure}[tbp]
	\centering
	\includegraphics[width=0.22\textwidth]{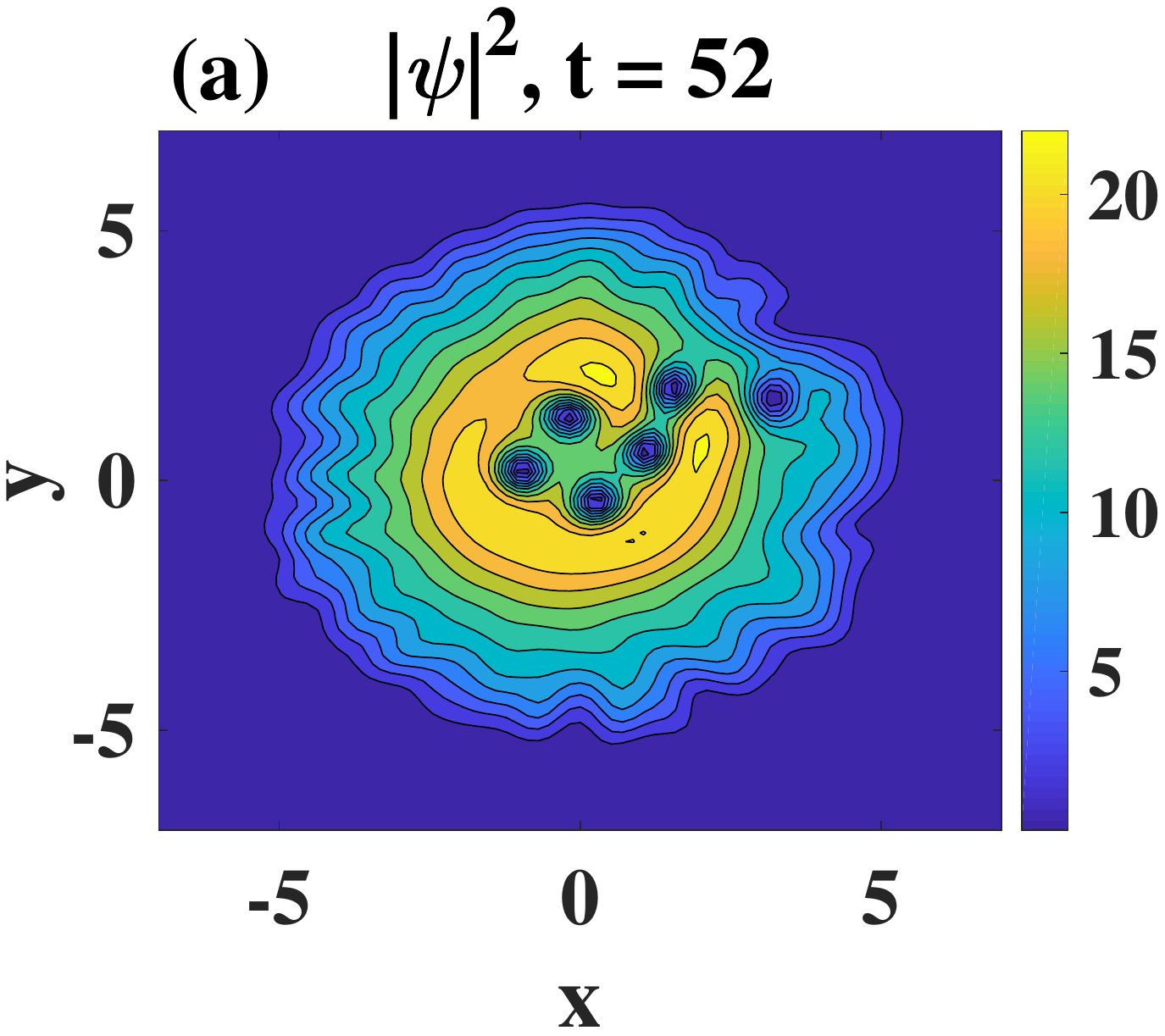}
	\includegraphics[width=0.22\textwidth]{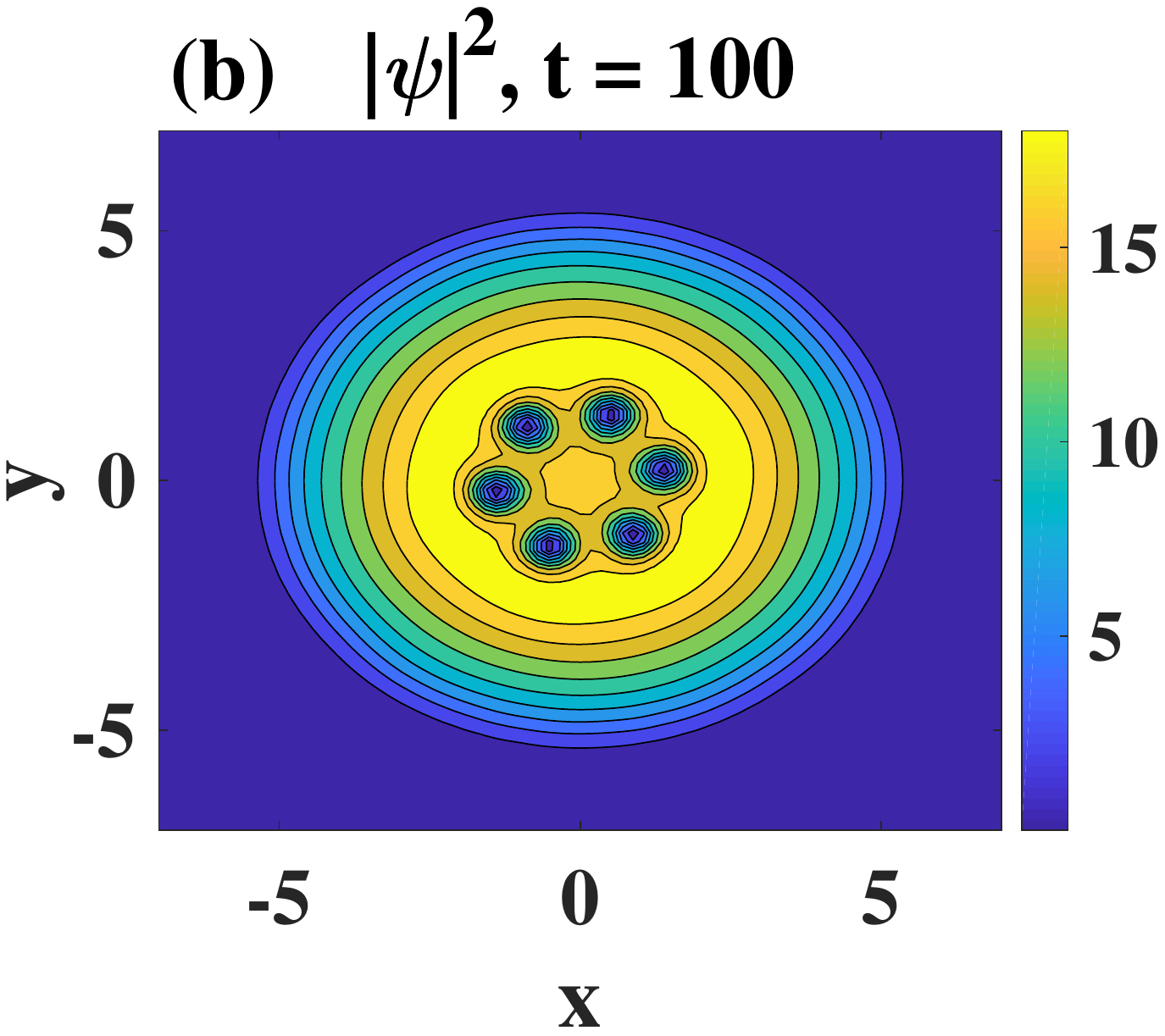}
	\\
	\vskip2mm
	\includegraphics[width=0.22\textwidth]{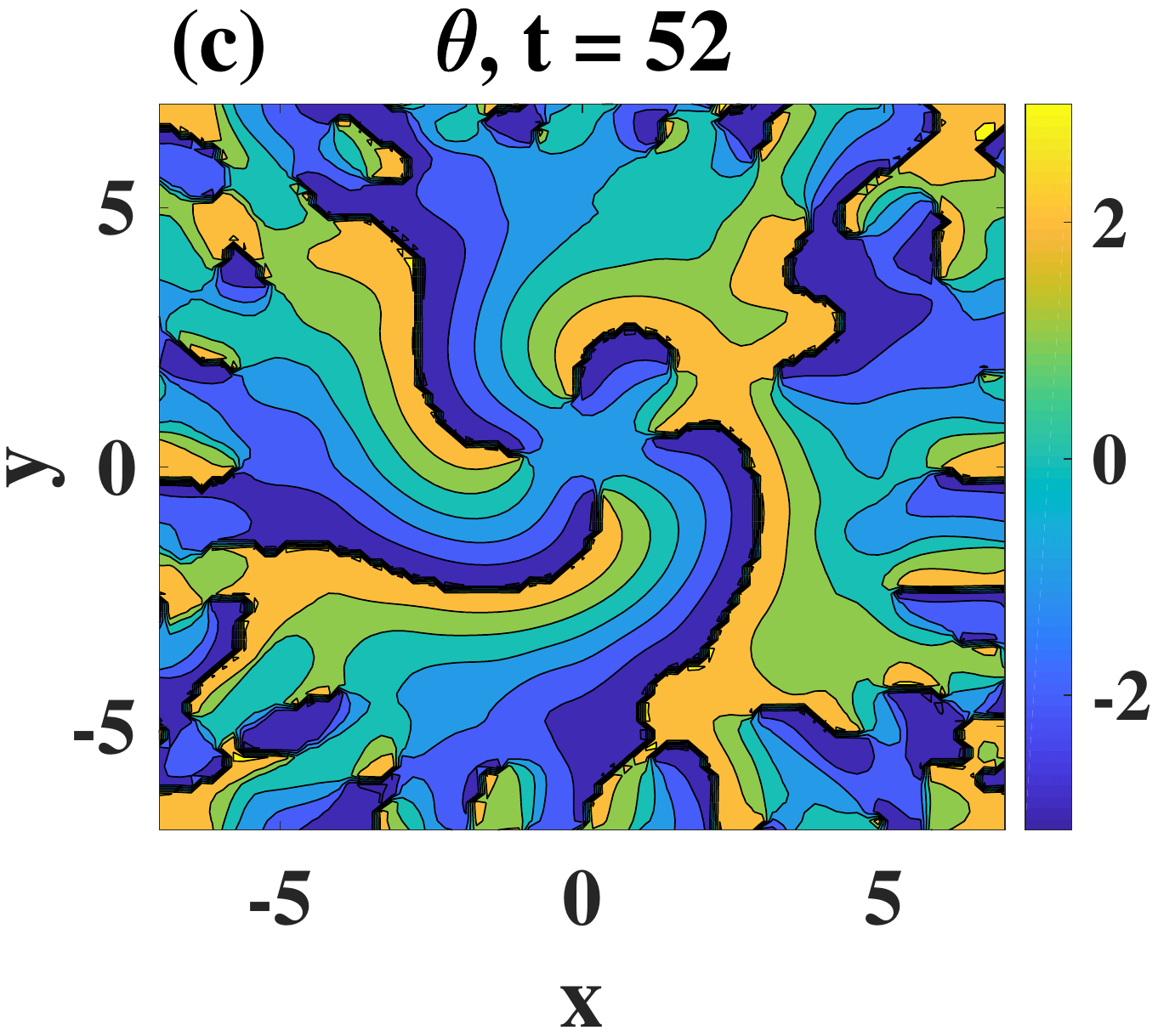}
	\includegraphics[width=0.22\textwidth]{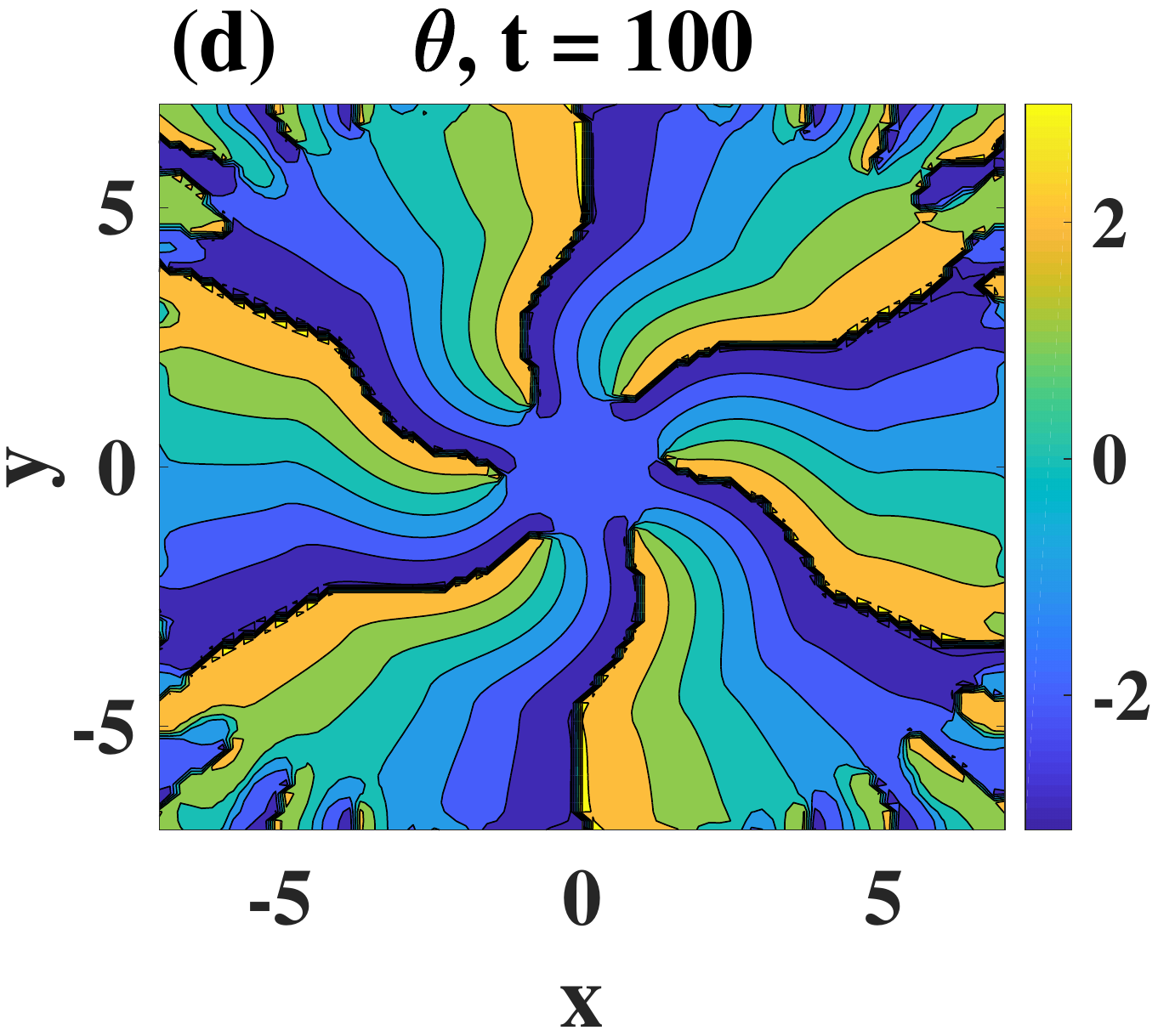}
	\\
	\vskip2mm
	\includegraphics[width=0.22\textwidth]{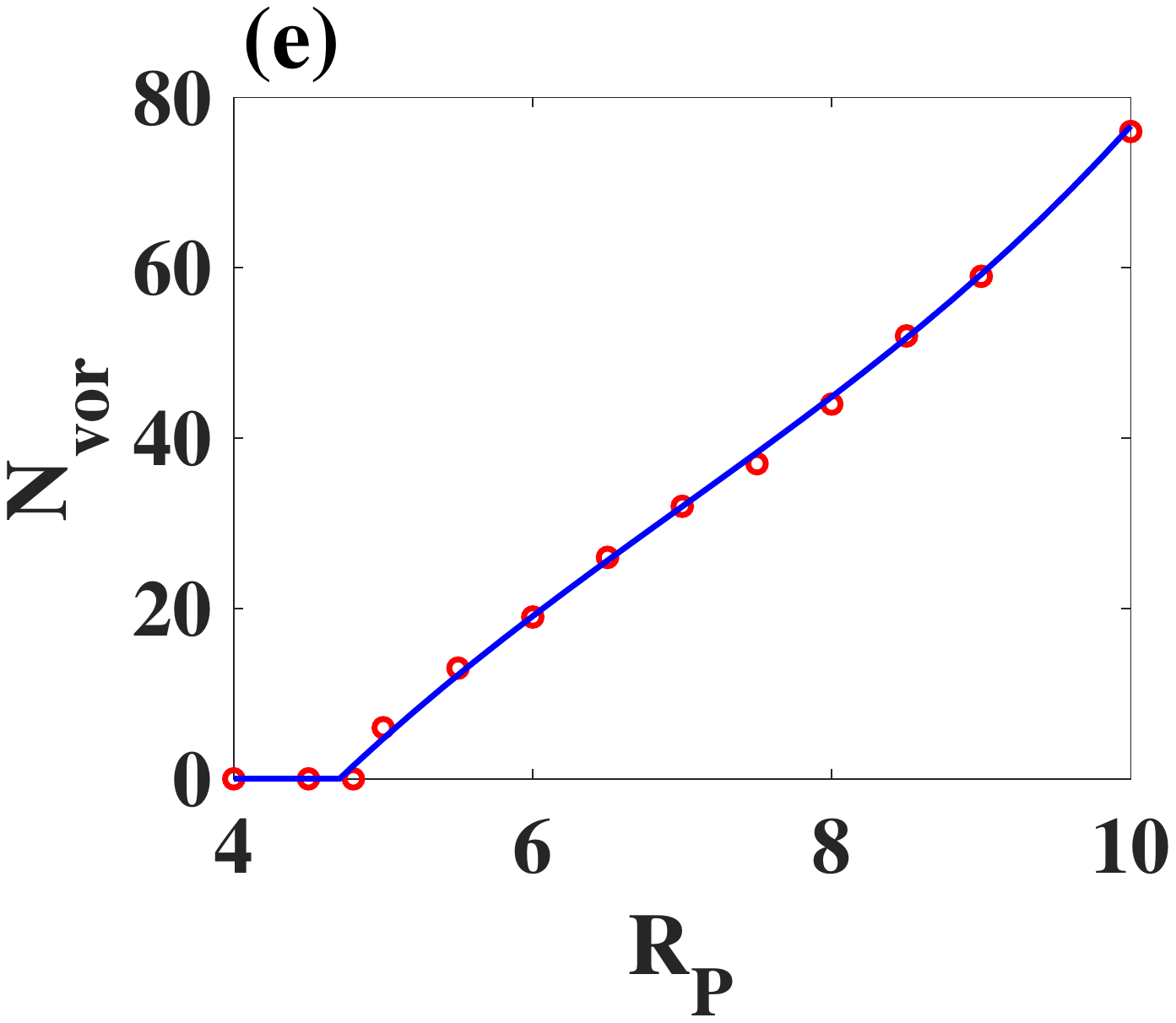}
	\includegraphics[width=0.22\textwidth]{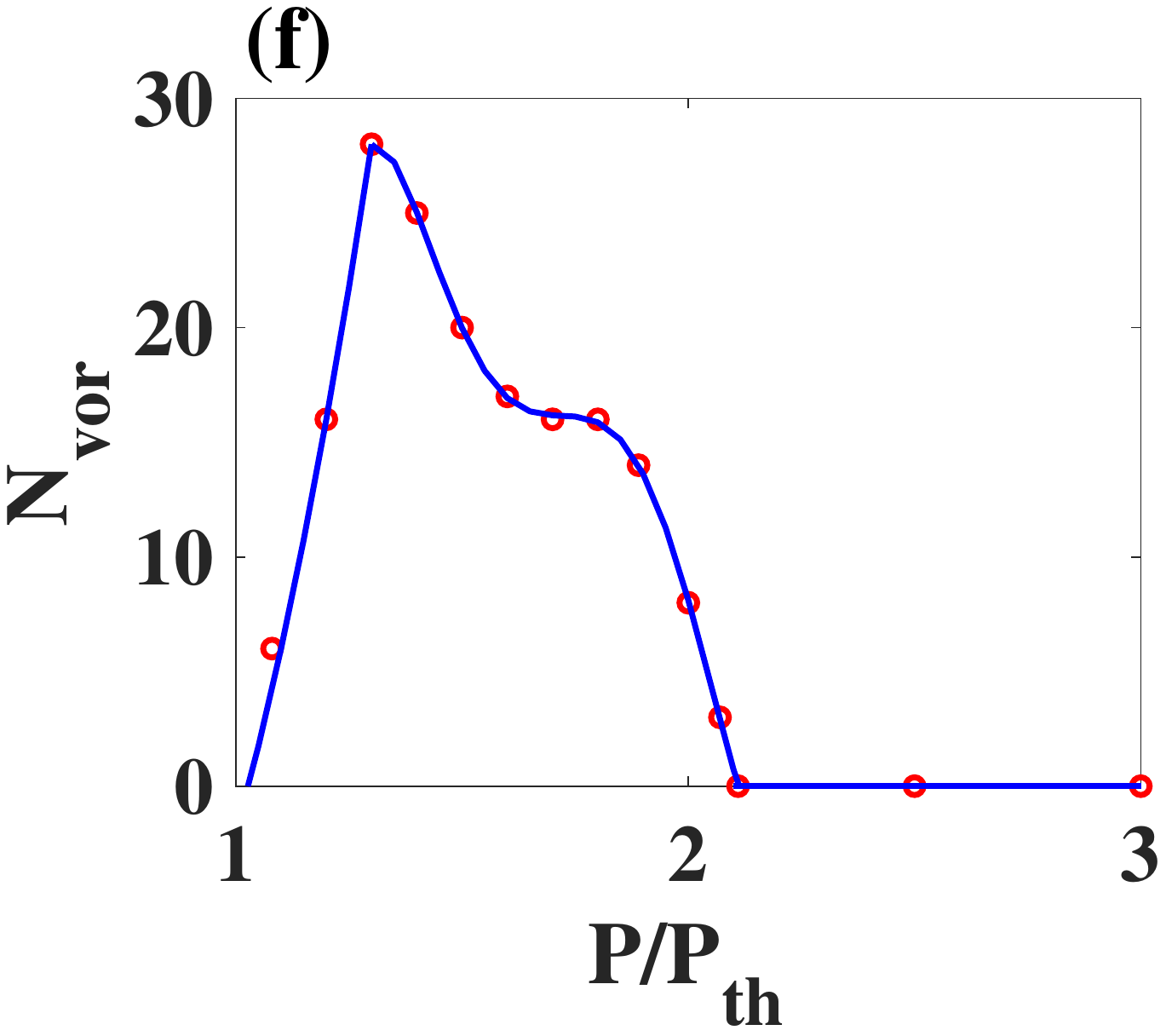}
	\caption{(Color online) Time evolution and steady-state solution of polariton condensate described by ODGPE~(\ref{eqn:ODGPE_rescaled}) when the adiabatic approximation is valid. The panels (a) and (b) give the time evolution of density of polariton condensate, while (c) and (d) show the corresponding phase distributions. Parameters are chosen as $g'_c=1$, $g'_R=2g'_c$, $\gamma'_c=110$, $\gamma'_R=100\gamma'_c$, $R'=0.51\gamma'_c$, $P'=1.08P'_{th}$, and $R_P=5$. These parameters are chosen in consistence with those used in Fig.~\ref{fig:case6-GP}. The panel (e) shows the number of vortices in the steady state with different radius of pumping spot $R_P$, and (f) displays that with different pumping $P'$, while other parameters are fixed as in (a)-(d). The red circles in (e) and (f) are the numerical results in the long-time limit, and the blue solid lines are guide for the eye.}
	\label{fig:case11}
\end{figure}

In order to show the equivalence of GGPE and ODGPE under adiabatic approximation, we demonstrate the emergence of vortices and vortex lattices in Fig.~\ref{fig:case11} by using ODGPE where adiabatic approximation is satisfied, with parameters can be reduced to those used in Fig.~\ref{fig:case6-GP}. We find that in the long-time steady state, the vortex lattice obtained by ODGPE [Figs.~\ref{fig:case11}(b) and \ref{fig:case11}(d)] is nearly the same as those by GGPE [Figs.~\ref{fig:case6-GP}(b) and \ref{fig:case6-GP}(d)]. This demonstrates the validity of GGPE when the adiabatic approximation is satisfied. On the other hand, the configurations of vortices at intermediate time scale as shown in Figs.~\ref{fig:case11}(a) and \ref{fig:case11}(c) are only qualitatively similar to the GGPE results of Figs.~\ref{fig:case6-GP}(a) and \ref{fig:case6-GP}(c). In fact, the time evolution of an open system towards its steady state is in general sensitively dependent on initial conditions and local perturbations. Thus, it is not unexpected that the time evolution of ODGPE and GGPE are consistent only qualitatively, even in the case of adiabatic approximation.

In Figs.~\ref{fig:case11}(e) and \ref{fig:case11}(f), we show respectively the number of vortices for different pumping radius $R_P$ and pumping power $P'$ in the vortex lattices. The dependence of number of vortices on the pumping radius $R_P$ and the pumping power $P$ are shown similar tendencies in Fig.~\ref{fig:case6-GP}. This verifies the equivalence of GGPE and ODGPE under adiabatic approximation as well.

\subsection{Spontaneous vortex lattices beyond the adiabatic approximation}
We then go beyond the adiabatic approximation, and numerically solve the ODGPE [Eq.~(\ref{eqn:ODGPE_rescaled})] under various conditions. Considering the relations of parameters of GGPE and ODGPE, as well as the Thomas-Fermi radius $R_{TF}$ which defines the minimal pumping radius to generate vortices and vortex lattices in polariton condensates, it is very subtle to choose parameters beyond the adiabatic approximation. In addition to choosing dampings $\gamma'_c$ and $\gamma'_R$ to define different regions beyond the adiabatic approximation, we also need to choose the pumping power $P'$, scattering rate $R'$ and interactions $g'_c$ and $g'_R$ properly to make the corresponding Thomas-Fermi radius $R_{TF}$ smaller than the pumping radius $R_P$. Only in this way we can observe spontaneous generation of vortices and vortex lattices in polariton condensates.

\begin{figure}[tbp]
	\centering
	\includegraphics[width=0.22\textwidth]{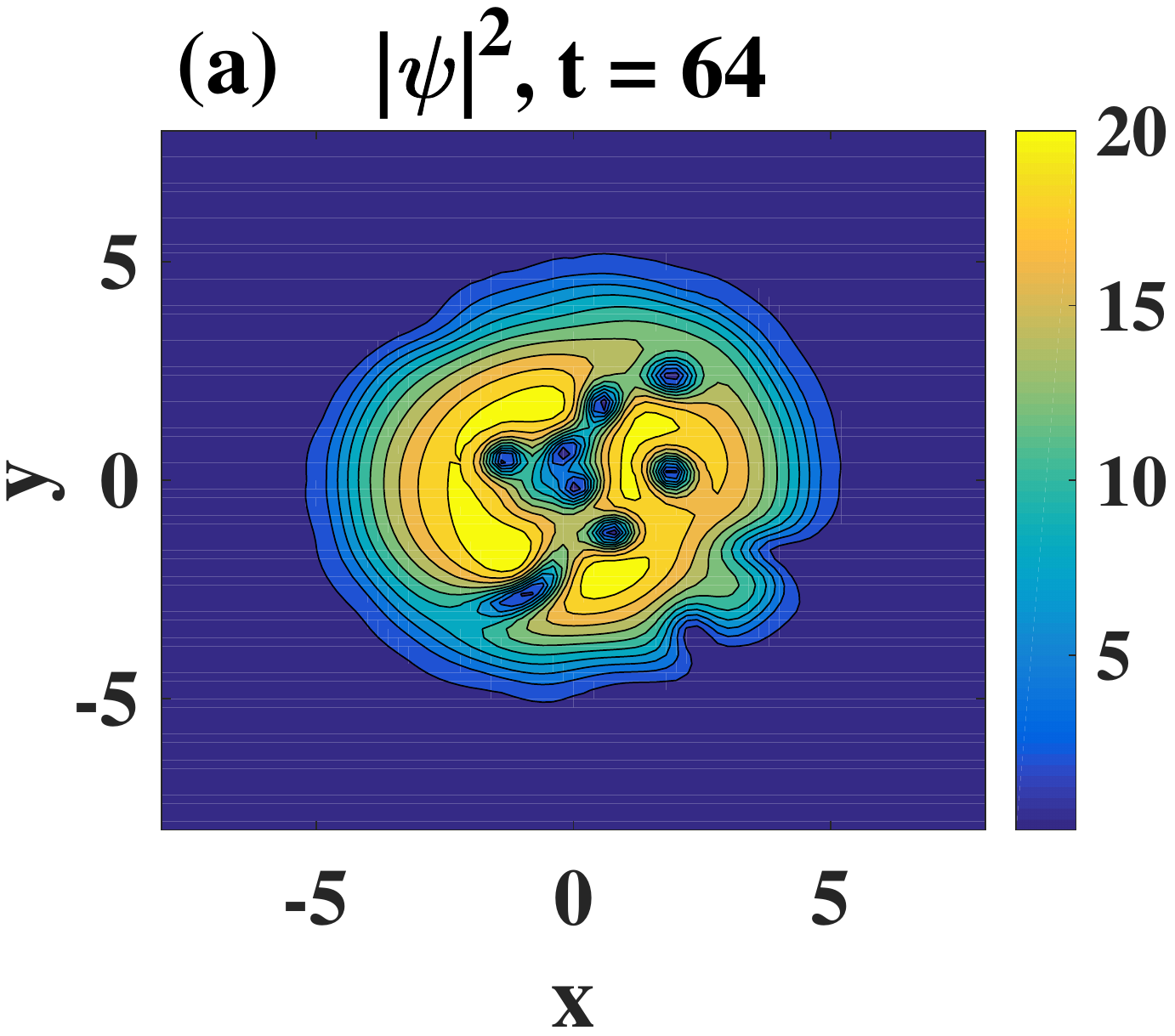}
	\includegraphics[width=0.22\textwidth]{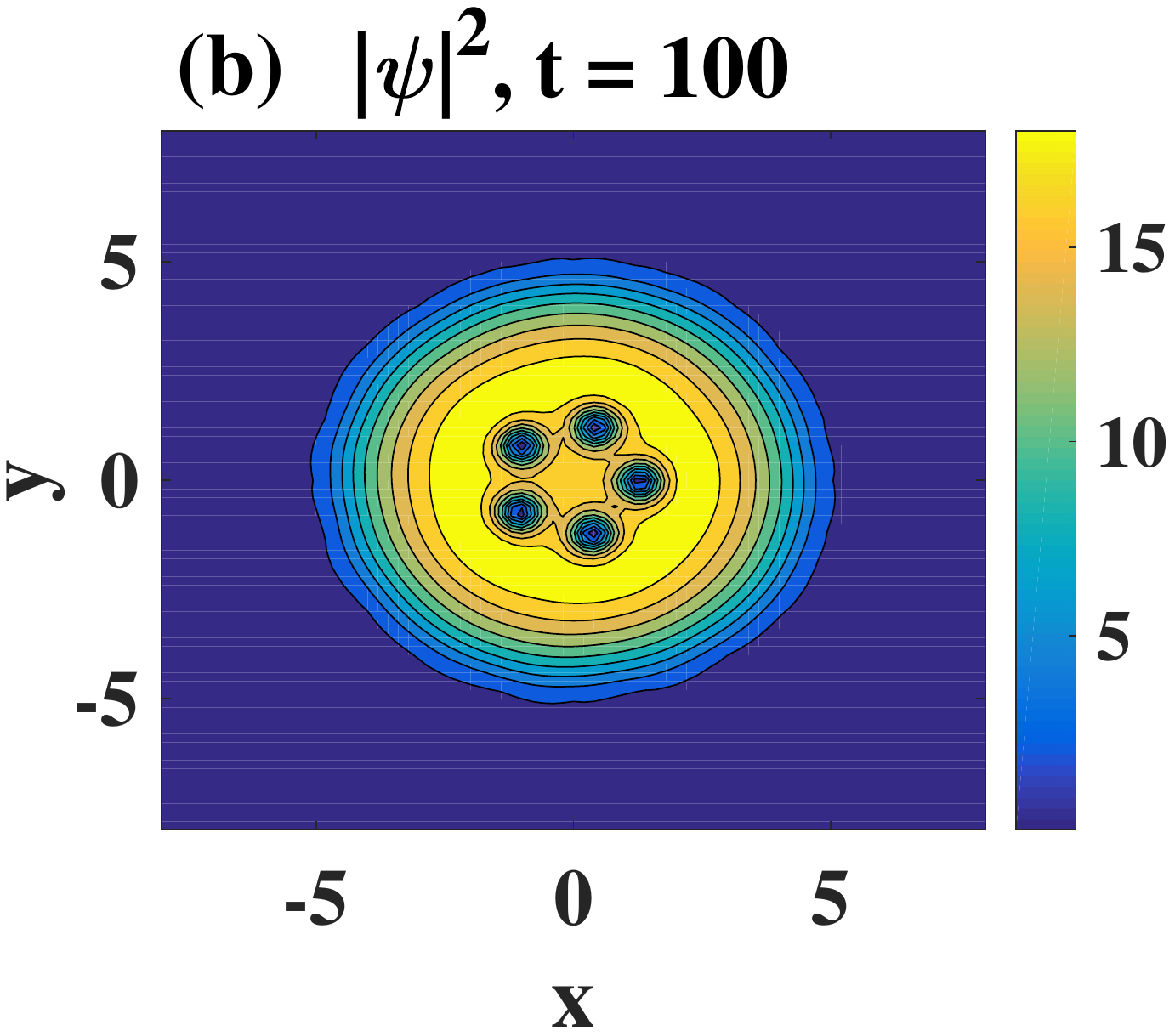}
	\\
	\vskip2mm
	\includegraphics[width=0.22\textwidth]{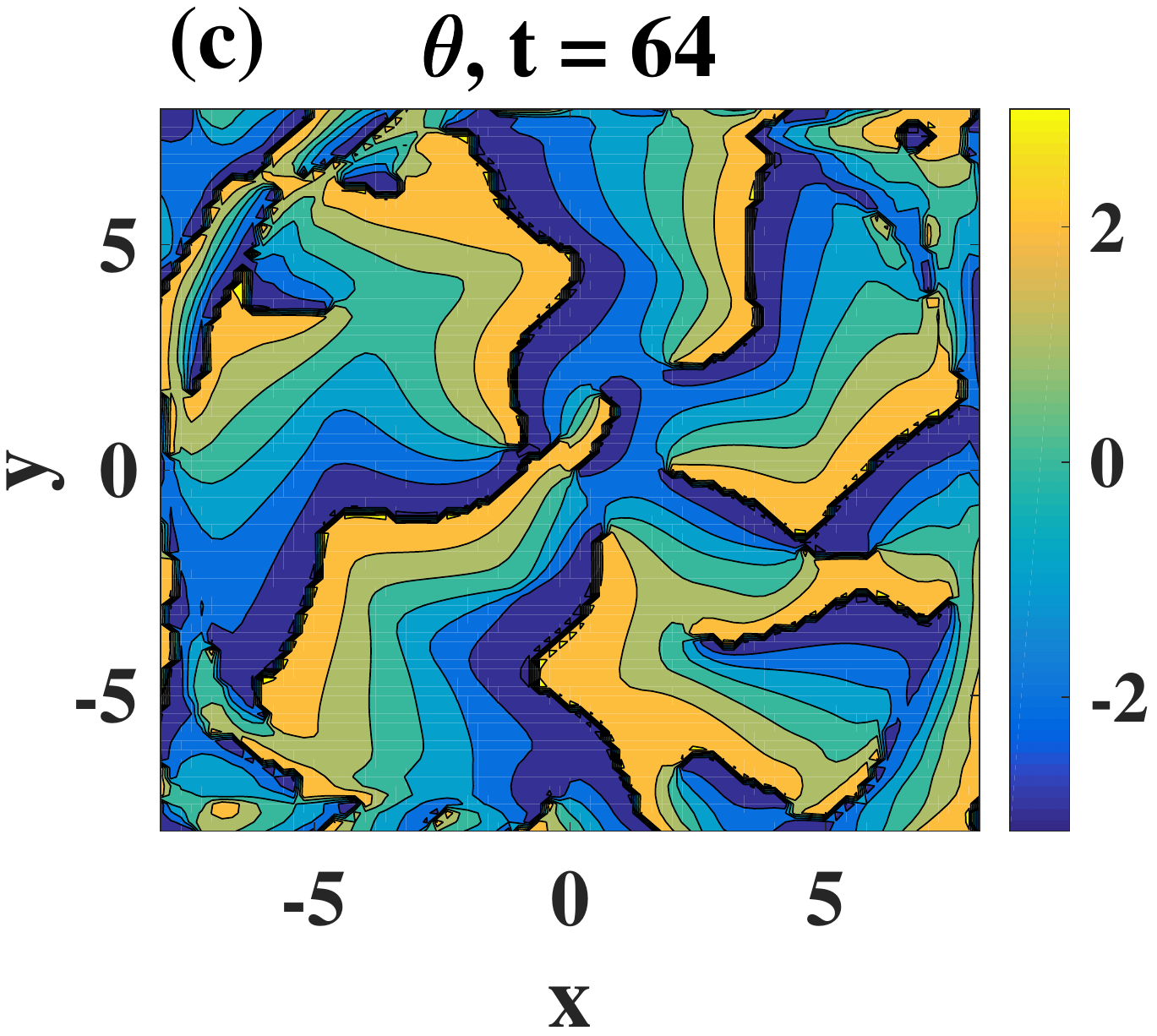}
	\includegraphics[width=0.22\textwidth]{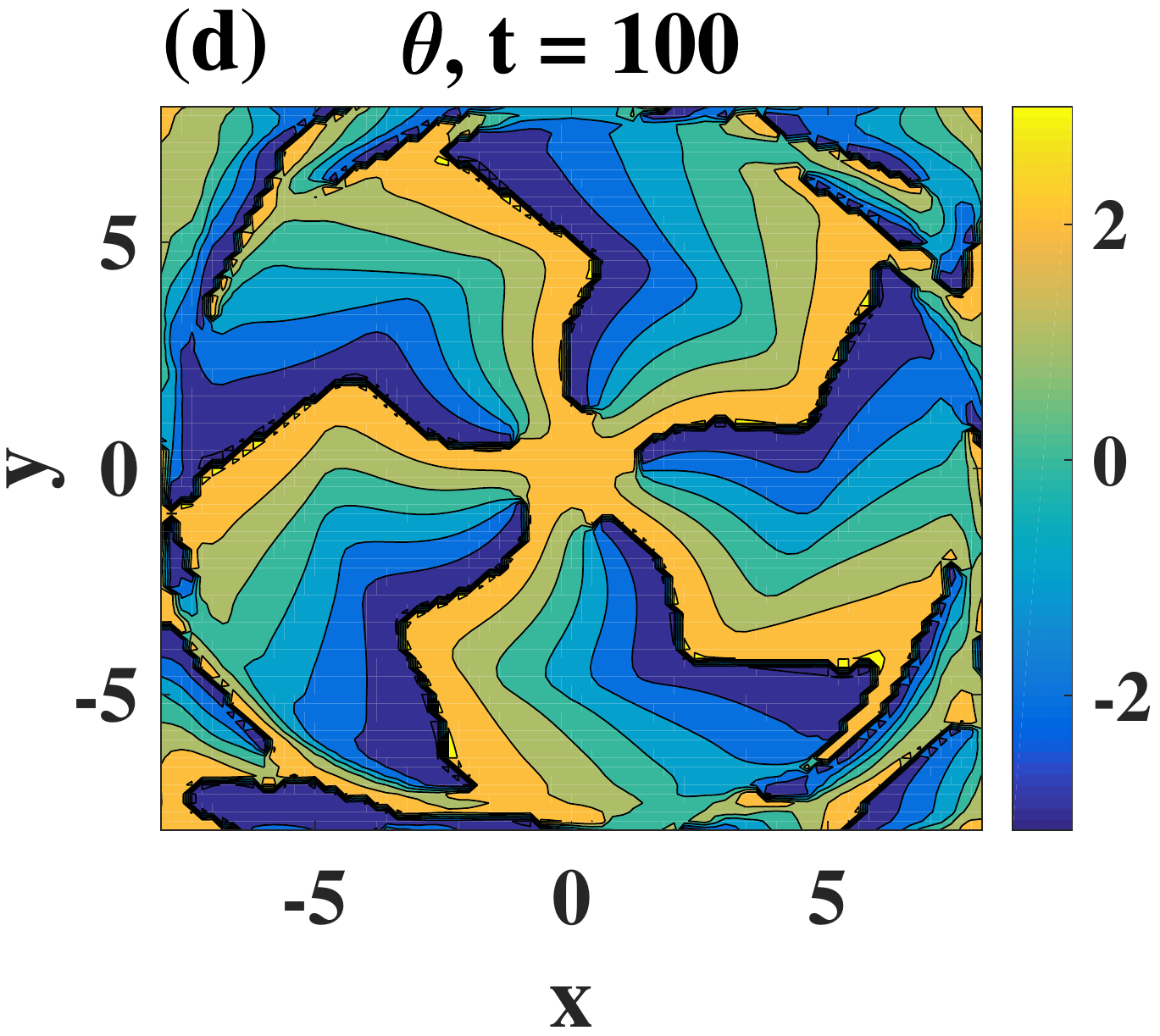}
	\\
	\vskip2mm
	\includegraphics[width=0.22\textwidth]{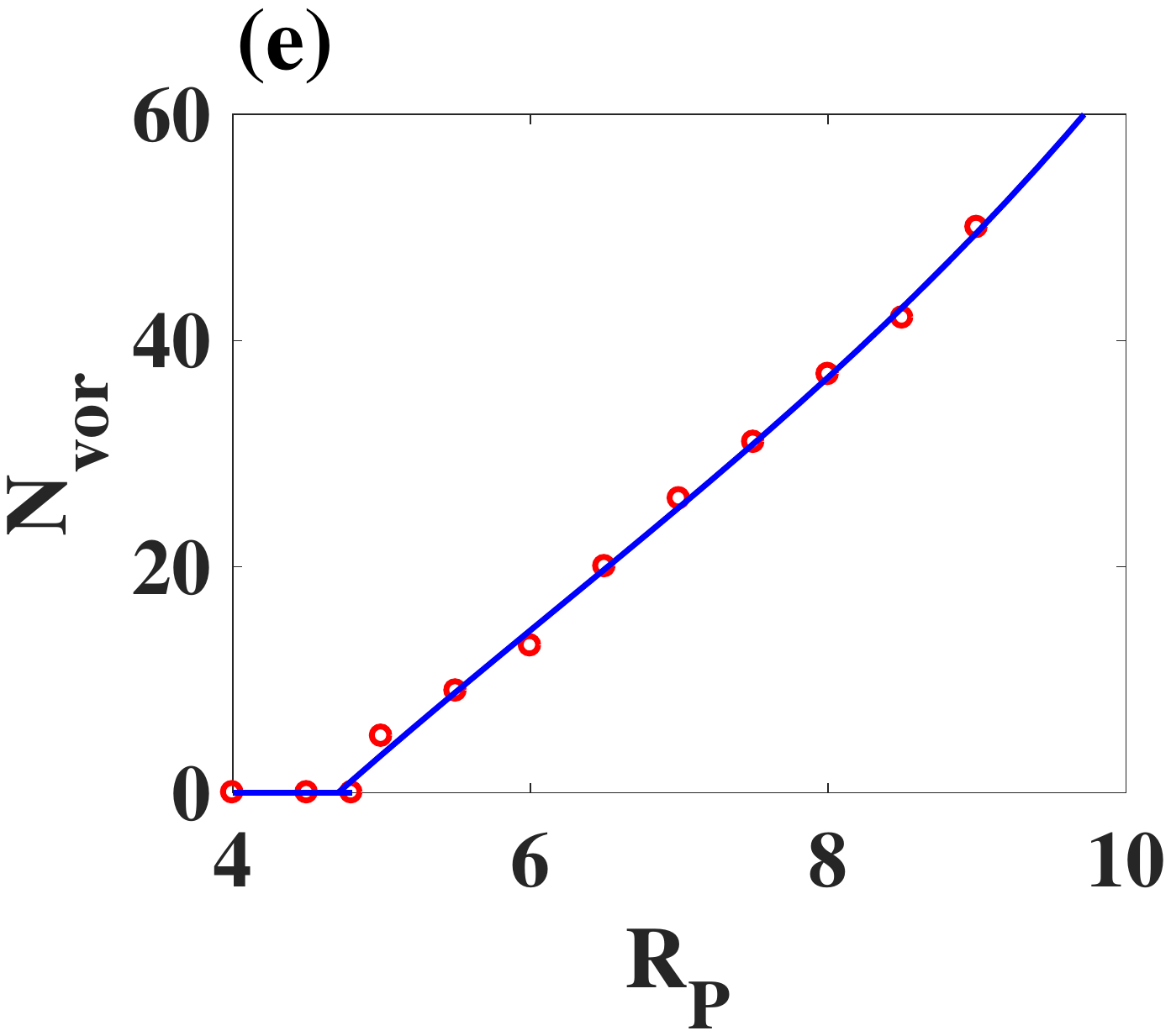}
	\includegraphics[width=0.22\textwidth]{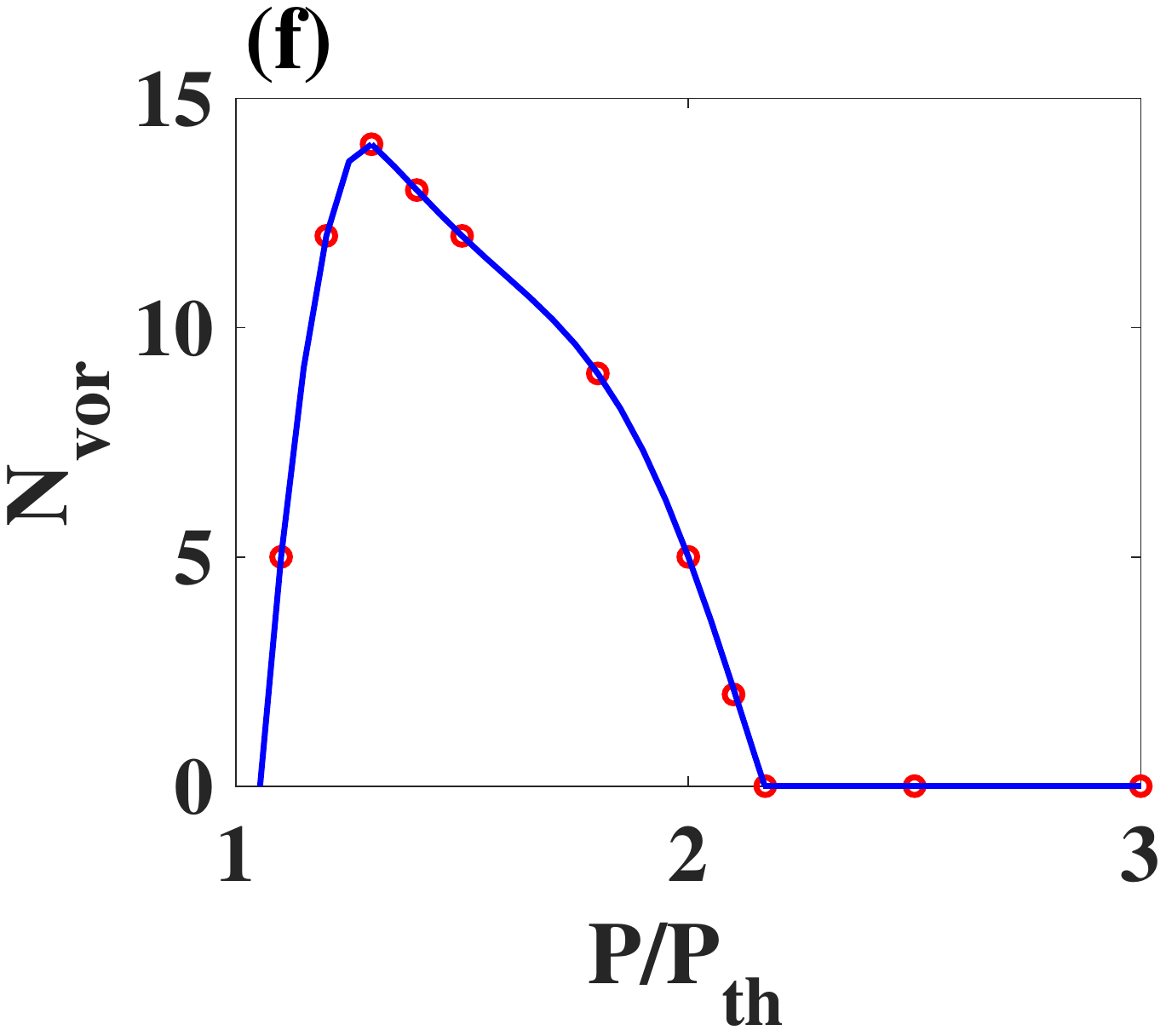}
	\caption{(Color online) Time evolution and steady-state solution of polariton condensate described by ODGPE~(\ref{eqn:ODGPE_rescaled}) where the adiabatic approximation partly breaks. The panels (a) and (b) give the time evolution of density of polariton condensate, while (c) and (d) show the corresponding phase distributions. Parameters are chosen as $g'_c=1$, $g'_R=2g'_c$, $\gamma'_c=8.8$, $\gamma'_R=10\gamma'_c$, $R'=0.68\gamma'_c$, $P'=2P'_{th}$, and $R_P=5$. The panel (e) shows the number of vortices in the steady state with different radius of pumping spot $R_P$, and (f) displays that with different pumping $P'$, while other parameters are fixed as in (a)-(d). The red circles in (e) and (f) are the numerical results in the long-time limit, and the blue solid lines are guide for the eye.}
	\label{fig:case15}
\end{figure}

Firstly, we consider the case where the adiabatic approximation is partly broken, i.e., the condition $\gamma'_R\gg P'R'/2\gamma'_c$ is unsatisfied and $\gamma'_R\gg\gamma'_c$ remains hold. 
In Fig.~\ref{fig:case15}, we present the numerical results for $\gamma'_R=P'R'/2\gamma'_c$ and $\gamma'_R = 10 \gamma'_c$, which are chosen to be consistent with typical experimental setups.~\cite{roumpos2011single}
The evolutions of the density distribution [Figs.~\ref{fig:case15}~(a) and \ref{fig:case15}(b)] and the phase distribution [Figs.~\ref{fig:case15}~(c) and \ref{fig:case15}(d)] show that vortex lattice can be stabilized. Besides, the dependence of number of vortices on the variations of the pumping laser radius $R_P$ [Fig.~\ref{fig:case15}(e)] and the pumping strength $\alpha$ [Fig.~\ref{fig:case15}(f)] are also qualitatively consistent with the case within adiabatic approximation as depicted in Fig.~\ref{fig:case6-GP} and Fig.~\ref{fig:case11}, although the absolute number of vortices is significantly reduced. Thus, we conclude that the breakdown of condition $\gamma'_R\gg P'R'/2\gamma'_c$ is not essential for the formation of vortex lattices.

\begin{figure}[tbp]
	\centering
	\includegraphics[width=0.22\textwidth]{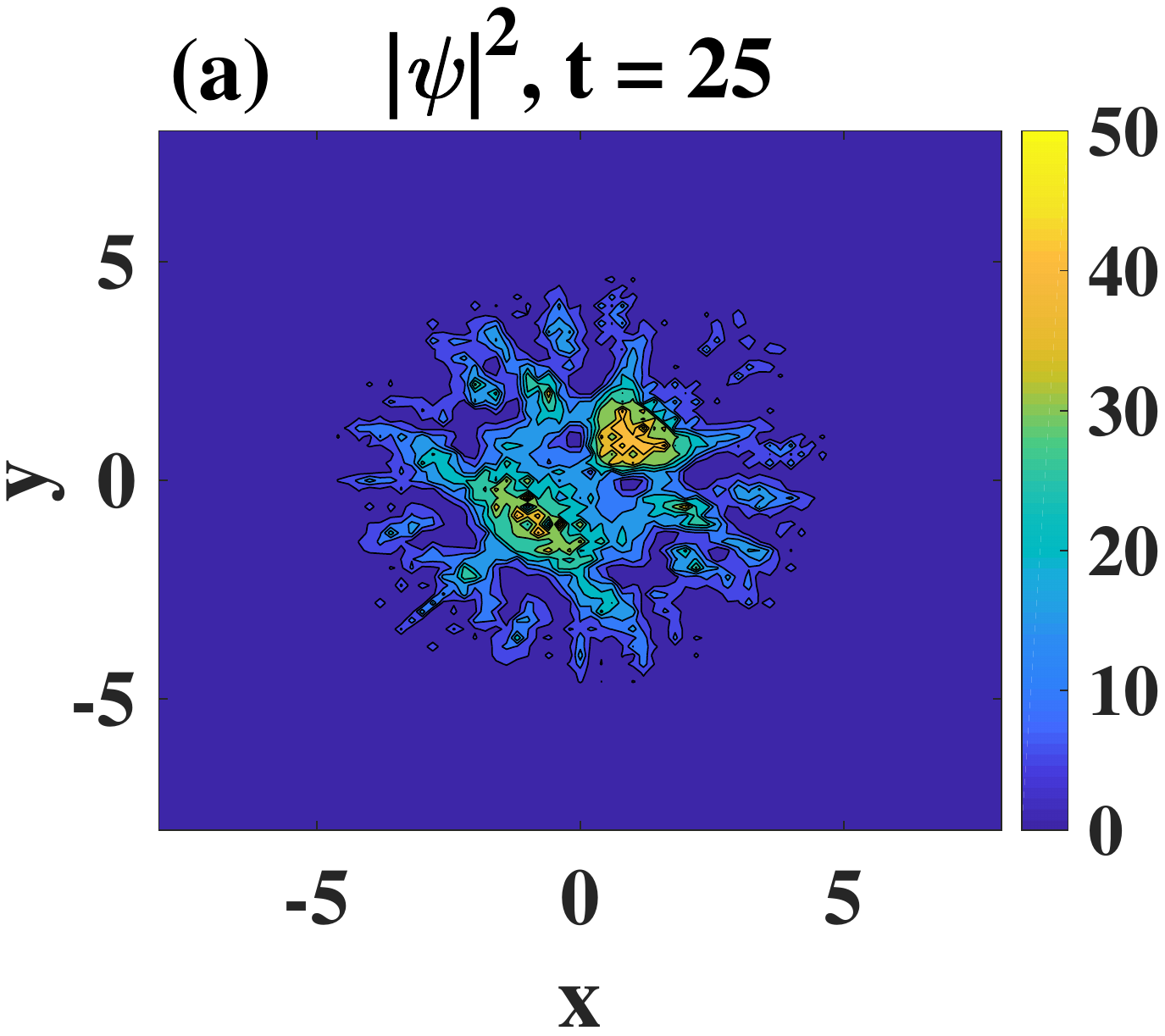}
	\includegraphics[width=0.22\textwidth]{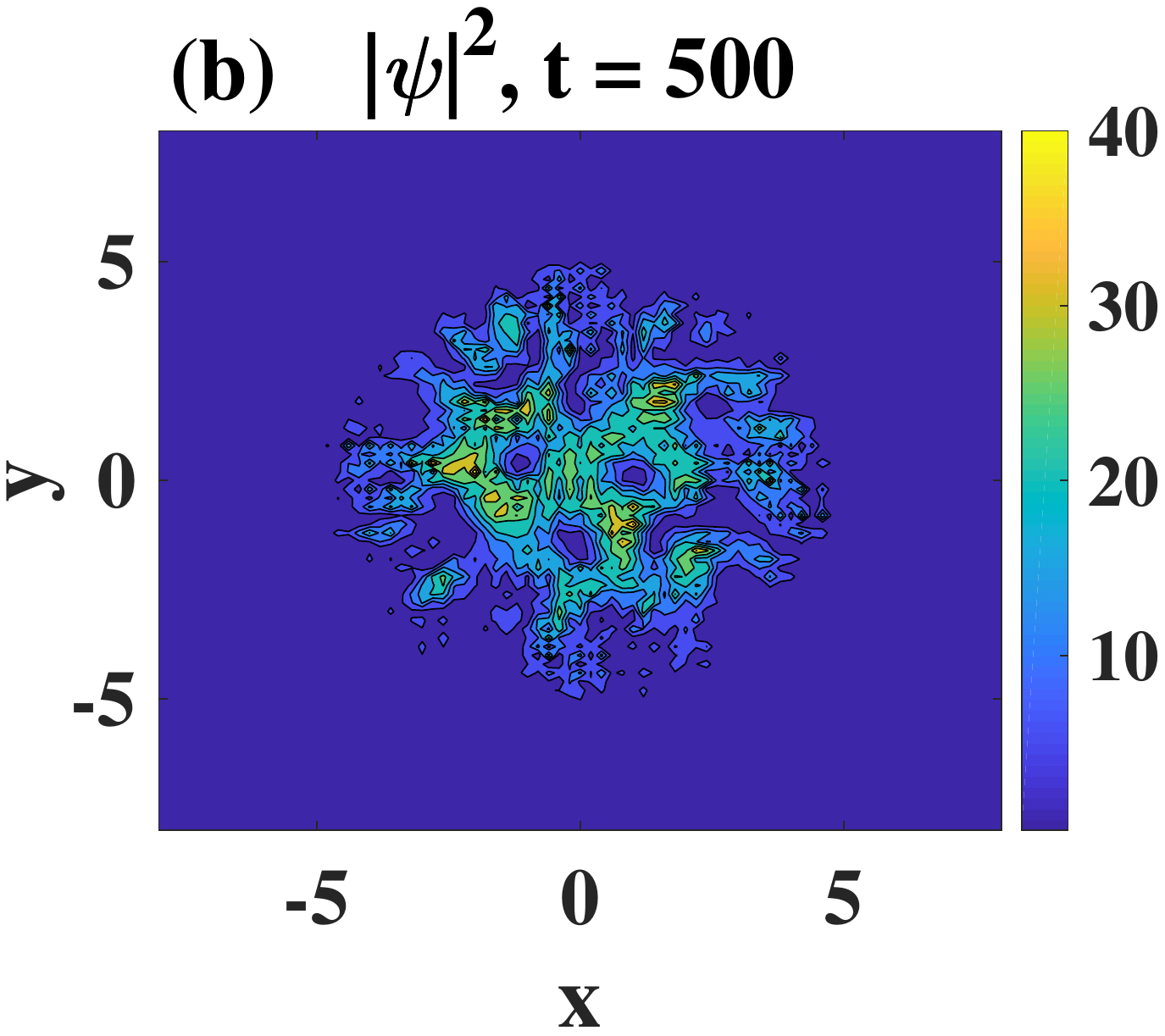}
	\\
	\vskip2mm
	\includegraphics[width=0.22\textwidth]{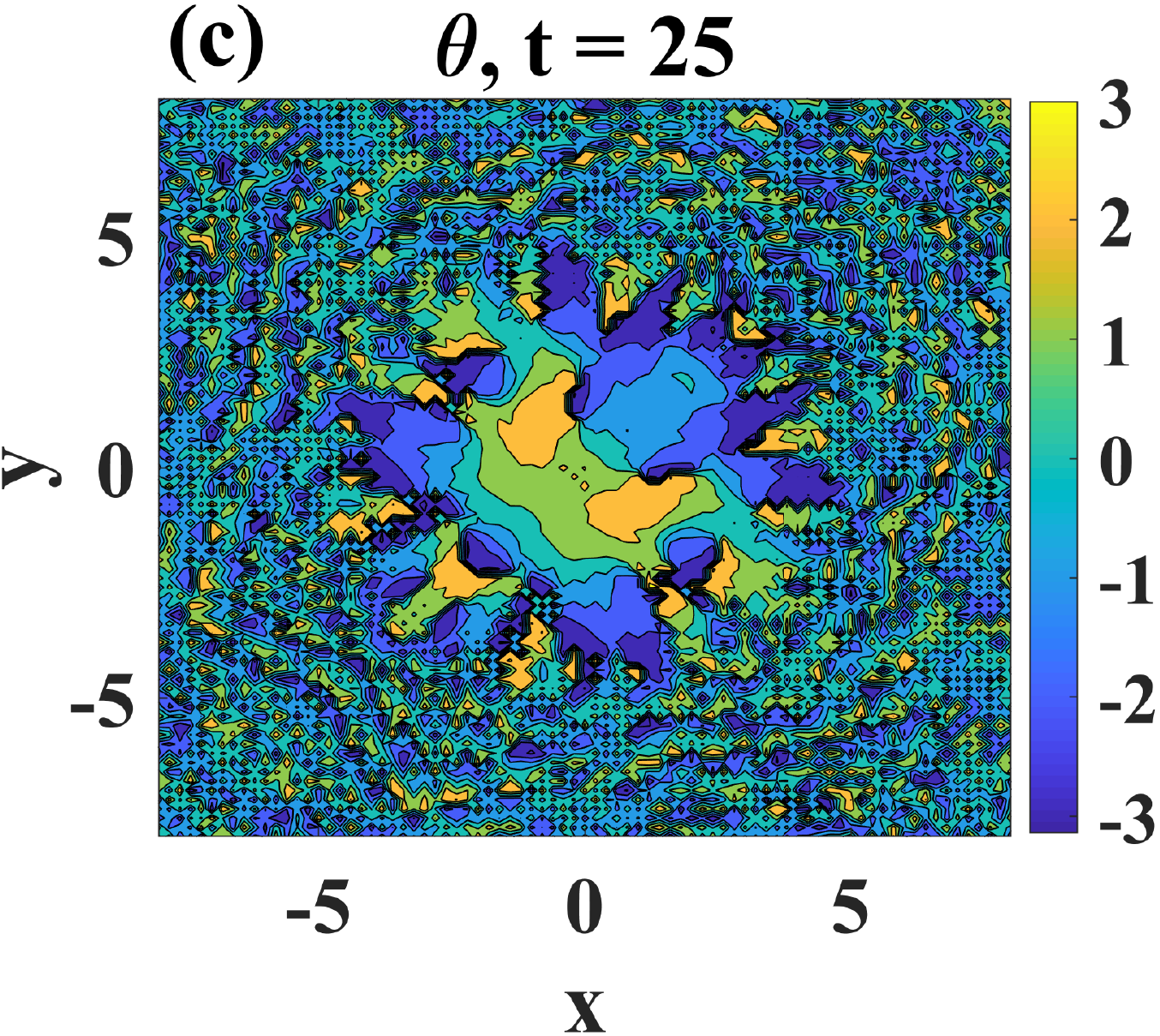}
	\includegraphics[width=0.22\textwidth]{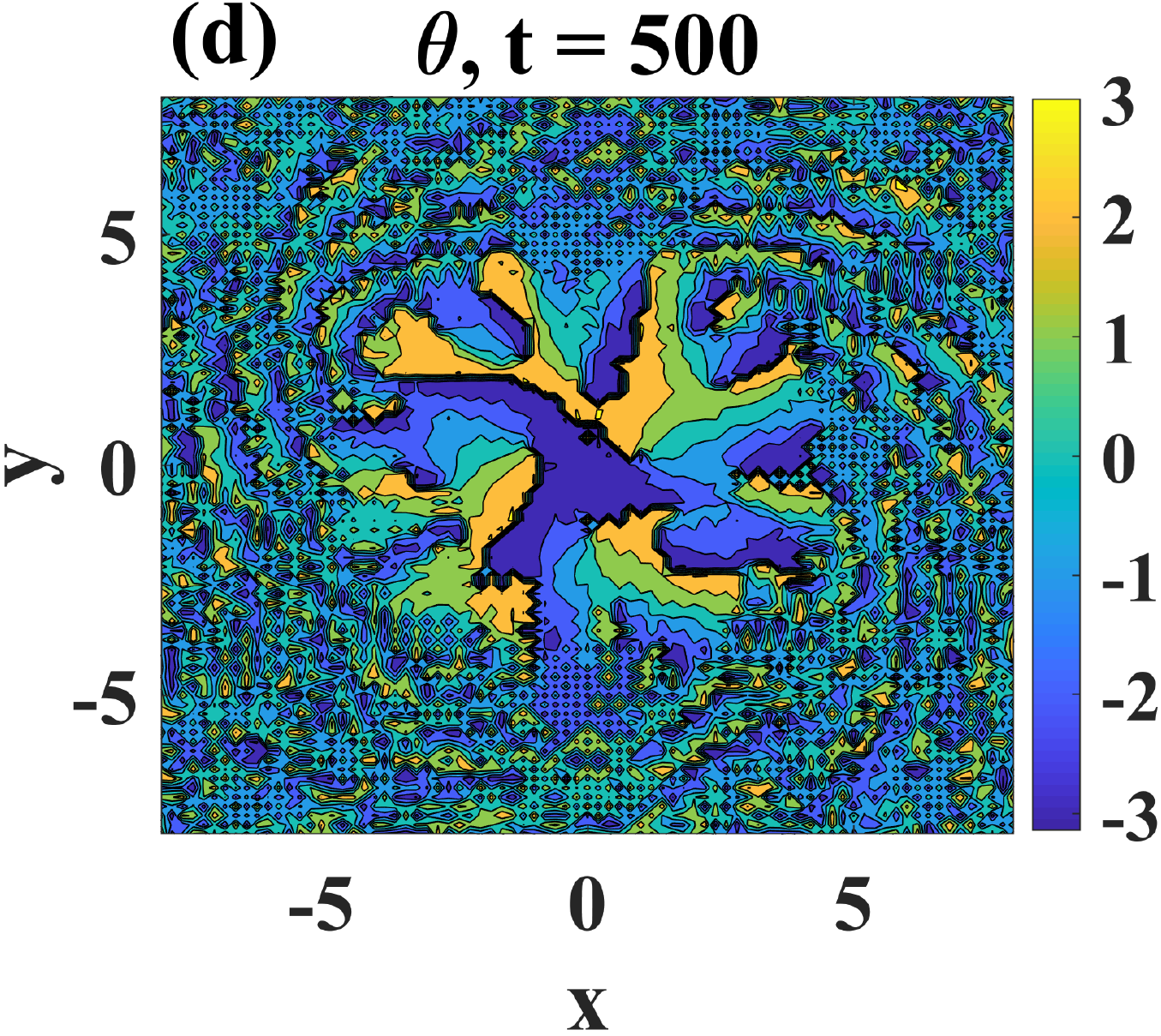}
	\caption{(Color online) Time evolution and steady-state solution of polariton condensate described by ODGPE~(\ref{eqn:ODGPE_rescaled}) where the adiabatic approximation completely breaks with $g'_c=1$, $g'_R=2g'_c$, $\gamma'_c=8.8$, $\gamma'_R=\gamma'_c$, and $R'=0.068\gamma'_c$. All panels display the same quantities as in Fig.~\ref{fig:case15}, where the laser radius and pumping power are chosen respectively as $R_P=5$ and $P'=2P'_{th}$.}
	\label{fig:case17}
\end{figure}

Next, we break the condition $\gamma'_R\gg\gamma'_c$ as well, and shown results for the case $\gamma'_R=\gamma'_c$ in Fig.~\ref{fig:case17}. In this scenario, we can still find spontaneously generated vortices from the density [Figs.~\ref{fig:case17}(a) and \ref{fig:case17}(b)] and phase [Fig.~\ref{fig:case17}(c) and \ref{fig:case17}(d)] distributions. However, a lattice structure can no longer be clearly recognized, indicating that the vortex lattice starts to melt and enter a liquid phase with number of vortices always changing with time. It is also shown that in this region, not only vortices but also anti-vortices are generated in the polariton condensates. The numbers of vortices and anti-vortices vary with time within the longest period of time of our numerical simulation, showing a prominent fluctuation effect in this regime.

\begin{figure}[tbp]
	\centering
	\includegraphics[width=0.22\textwidth]{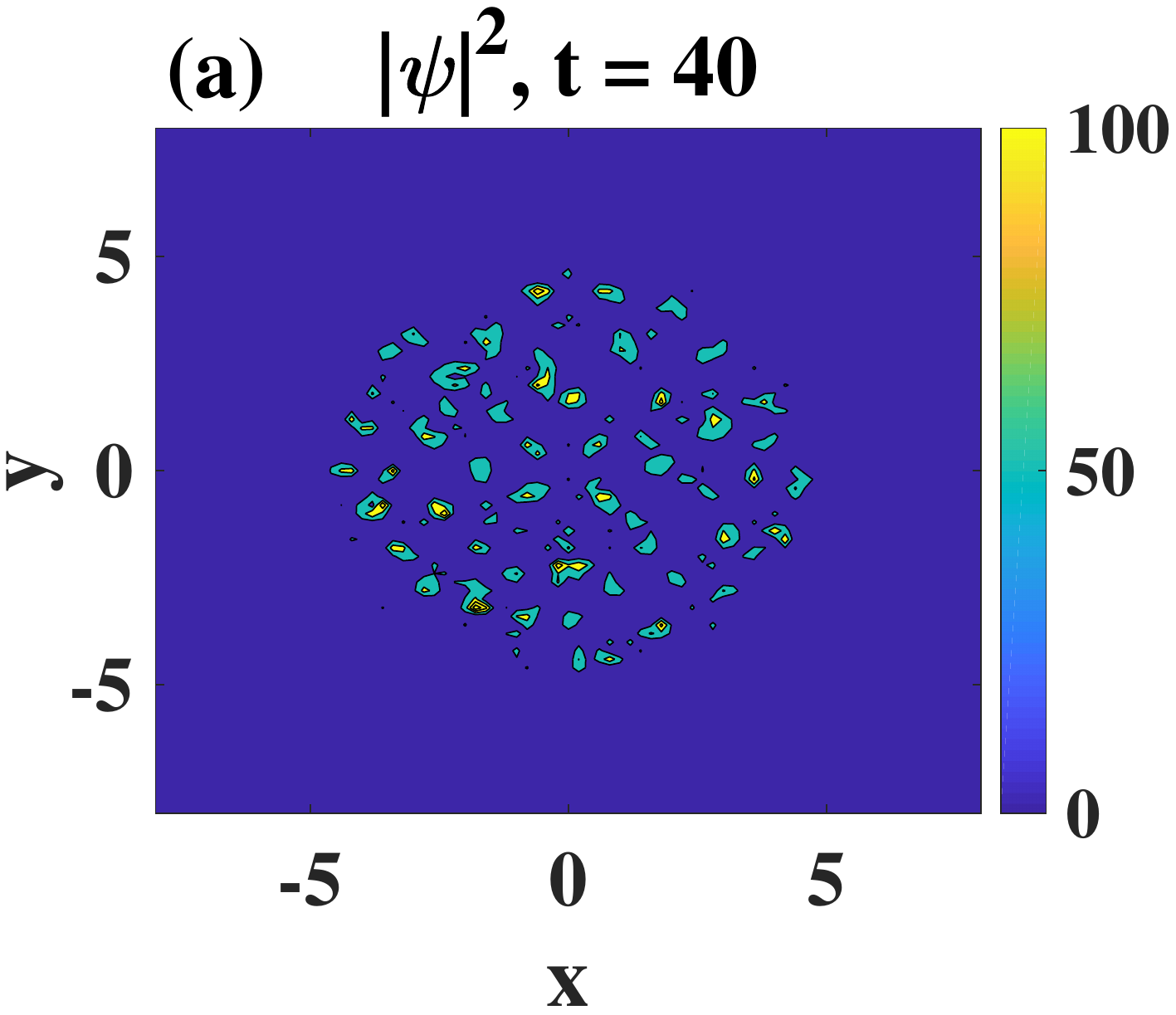}
	\includegraphics[width=0.22\textwidth]{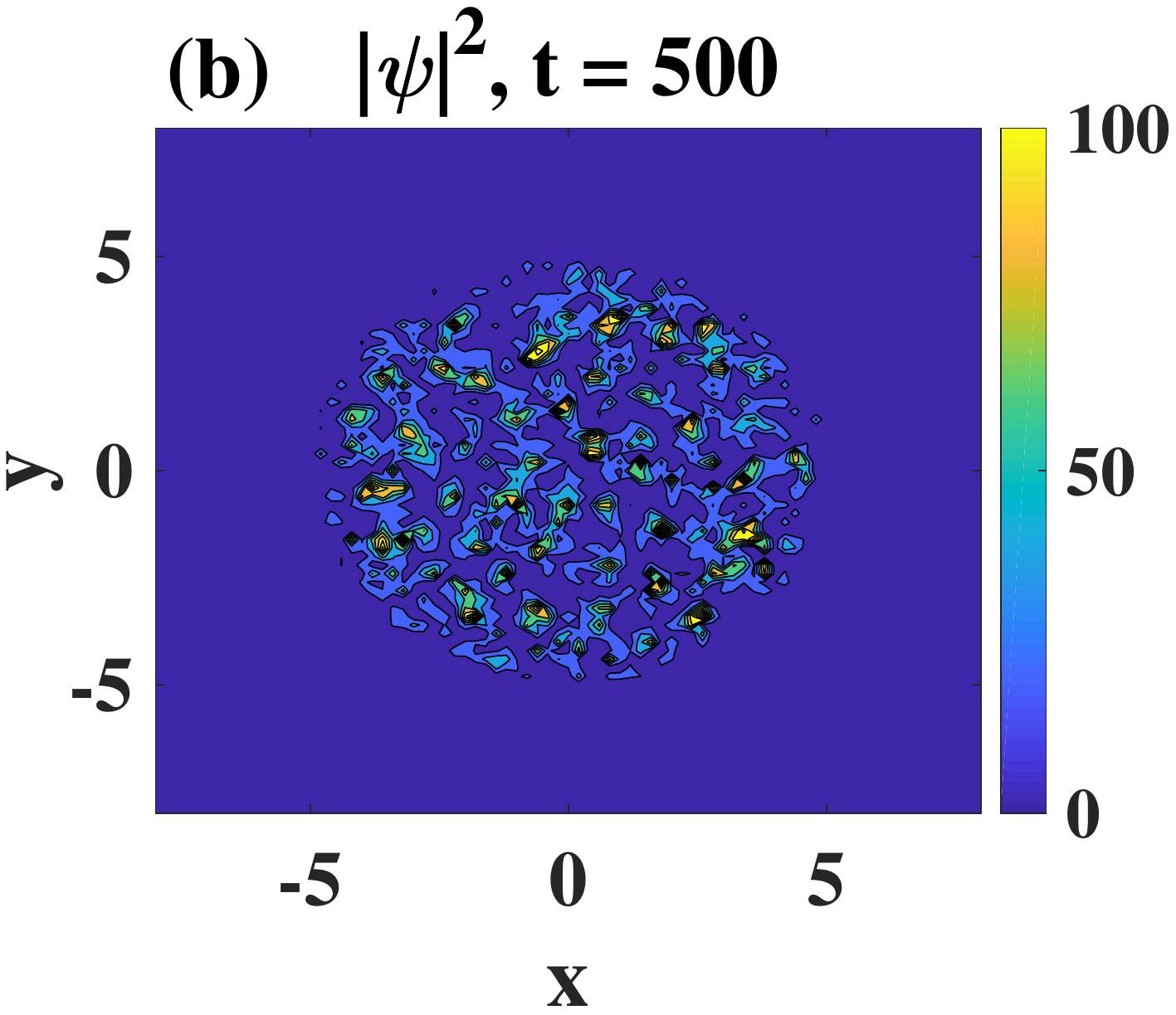}
	\\
	\vskip2mm
	\includegraphics[width=0.22\textwidth]{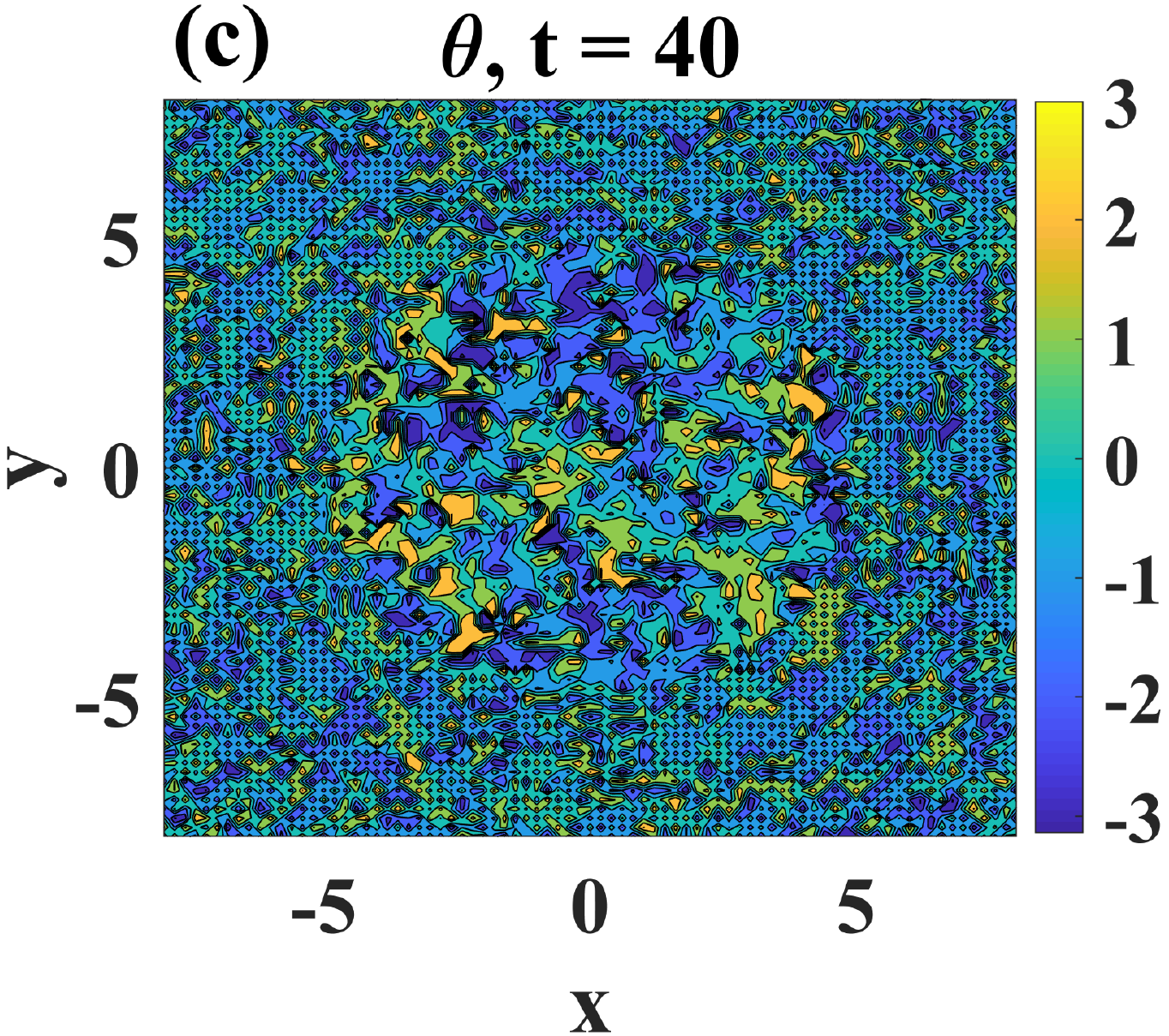}
	\includegraphics[width=0.22\textwidth]{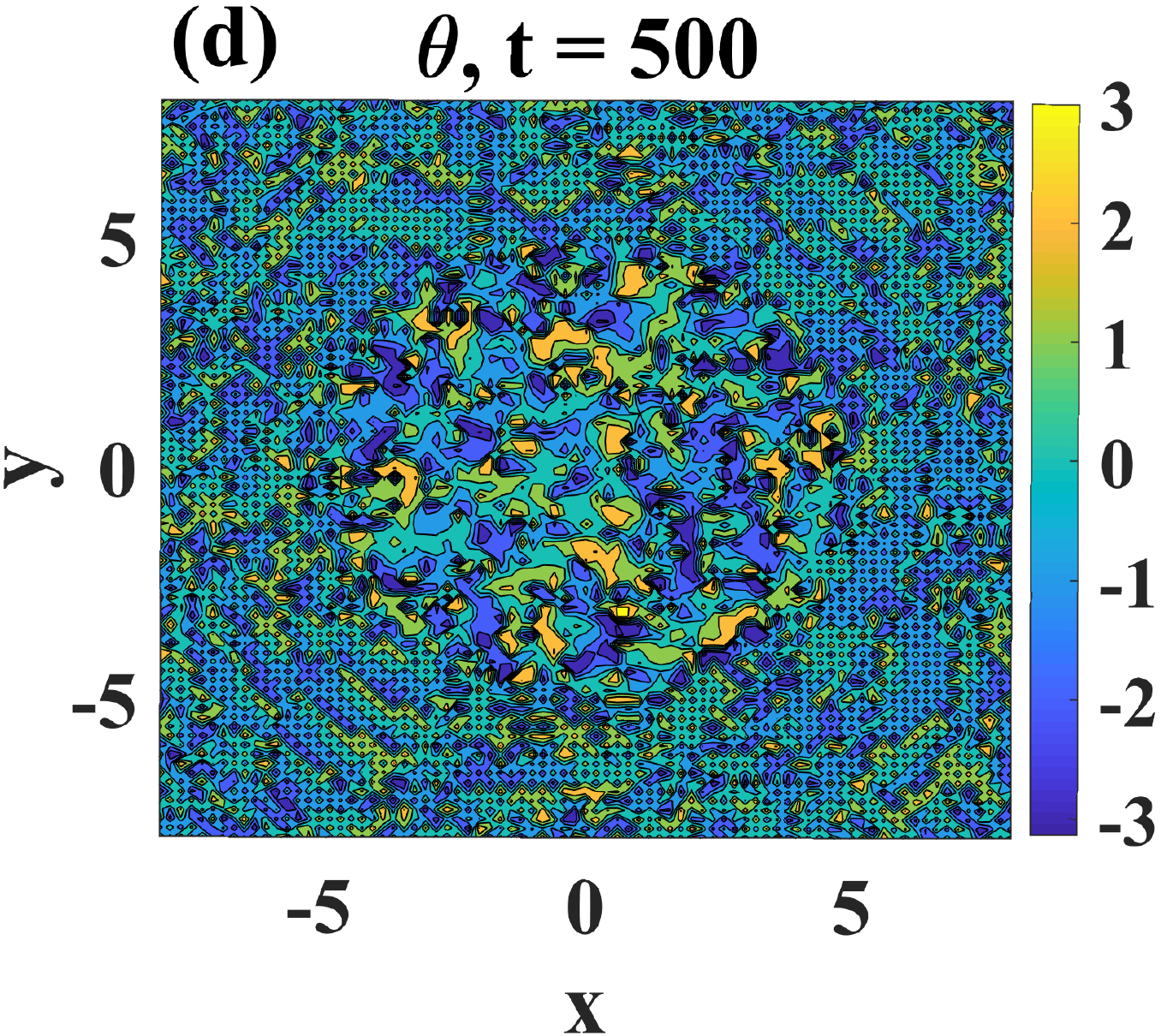}
	\caption{(Color online) Time evolution of polariton condensate described by ODGPE~(\ref{eqn:ODGPE_rescaled}) in the limit of $\gamma'_R\ll\gamma'_c$, with $g'_c=1$, $g'_R=2g'_c$, $\gamma'_c=1.26$, $\gamma'_R=0.1\gamma'_c$, and $R'=0.012\gamma'_c$. All panels display the same quantities as in Fig.~\ref{fig:case15}, where the laser radius and pumping power are chosen respectively as $R_P=5$ and $P'=8P'_{th}$.}
	\label{fig:case19}
\end{figure}

We can further reduce the decay rate of reservoir to reach the opposite limit of the adiabatic approximation $\gamma'_R\ll\gamma'_c$. 
The results for $\gamma'_R = 0.1 \gamma'_c$ are shown in Fig.~\ref{fig:case19}, with parameters are chosen in consistence with Ref.~\cite{takemura2017spin}. In this limiting case of adiabatic approximation broken, we find that not only the vortex lattices disappear, but the polariton condensates decompose as well. This observation can be understood by reminding that the effective interaction in GGPE $g=g'_c-g'_RP'R'/{\gamma'}_R^2$ depends on the reservoir decay rate, hence can become attractive with small enough $\gamma'_R$. Although the formalism of GGPE is not valid in this scenario, it can still be expected that a strong enough attractive interaction tend to destroy any steady Bose-Einstein condensates.

\begin{figure}[tbp]
	\centering
	\includegraphics[width=0.22\textwidth]{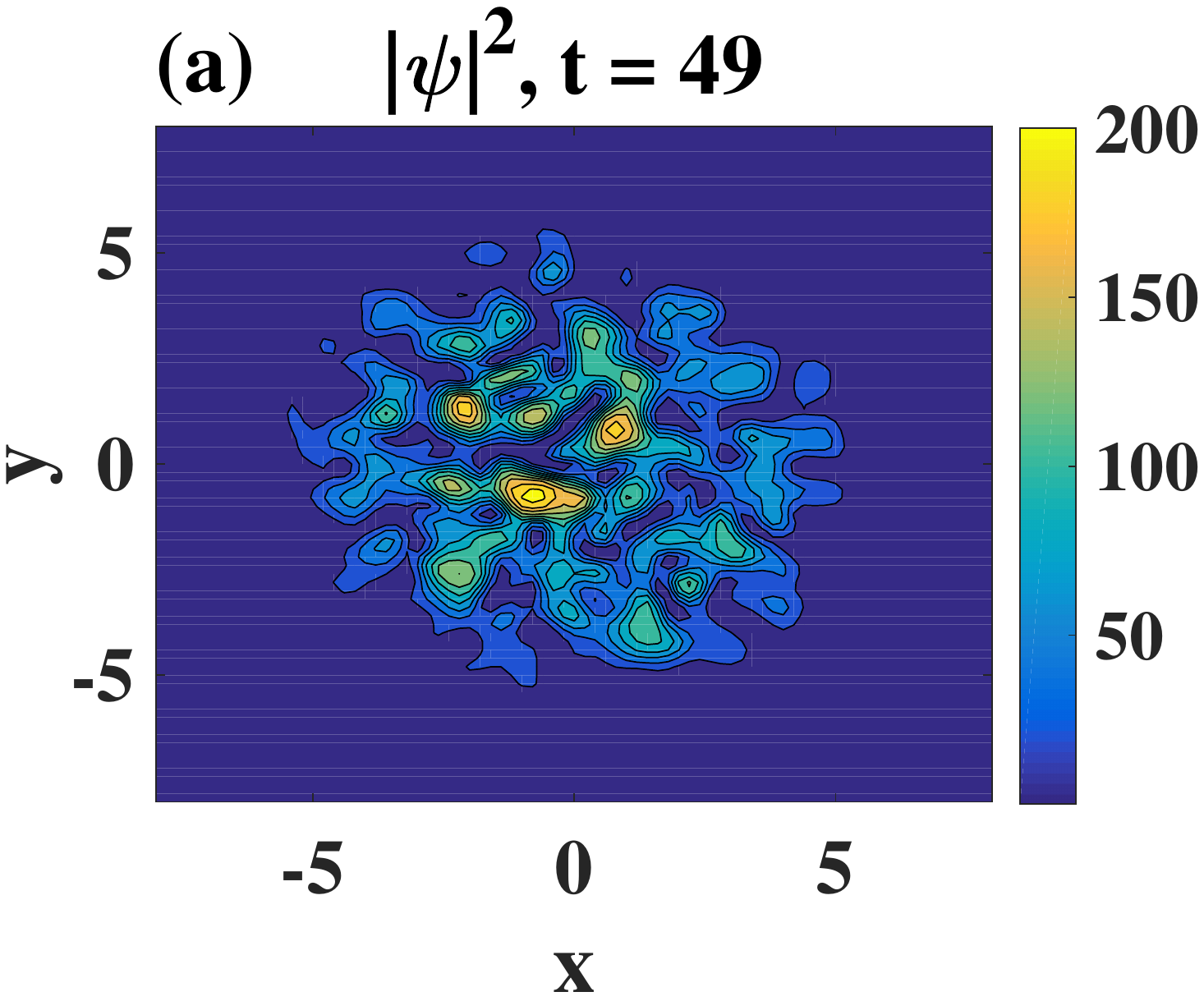}
	\includegraphics[width=0.22\textwidth]{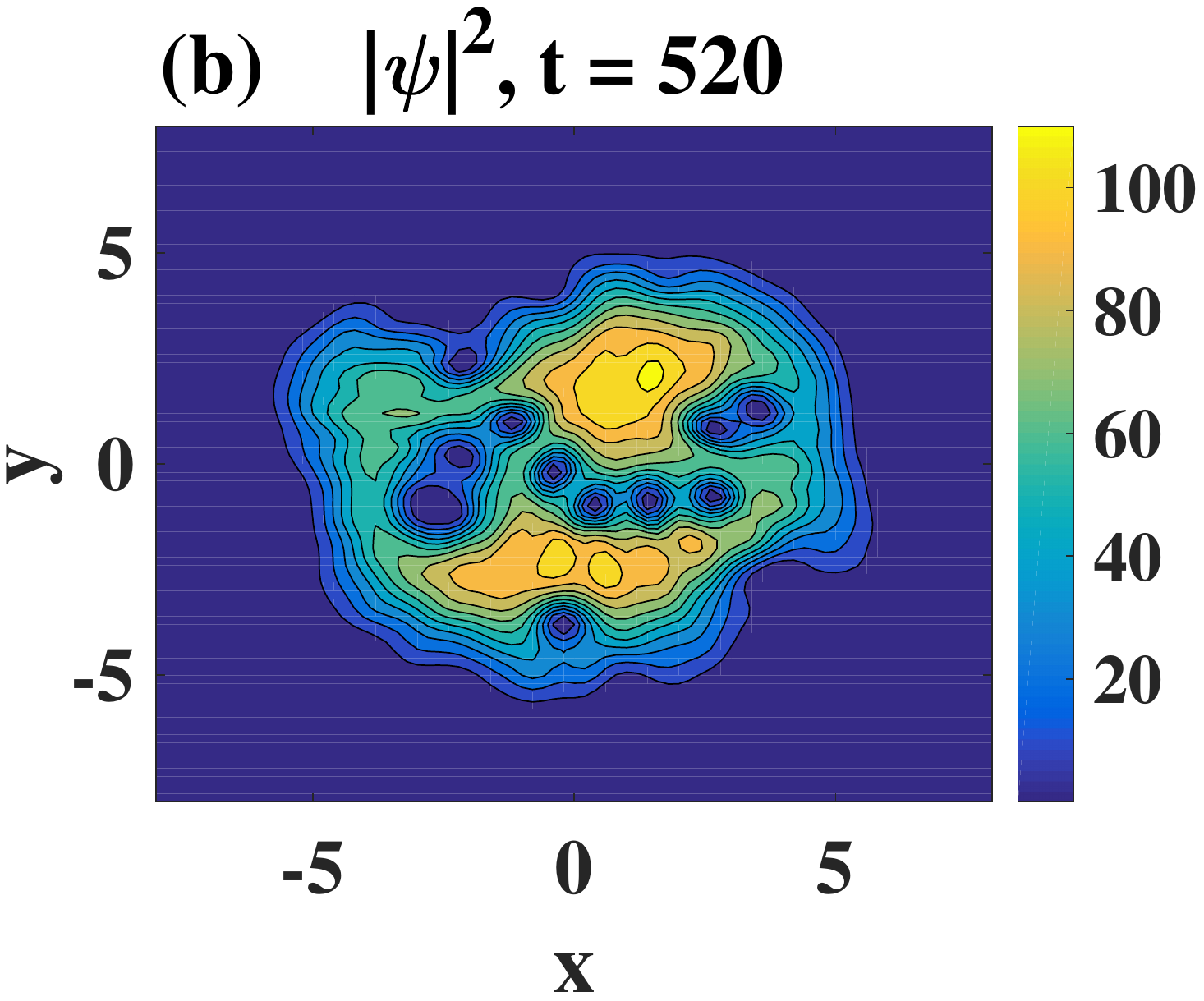}
	\\
	\vskip2mm
	\includegraphics[width=0.22\textwidth]{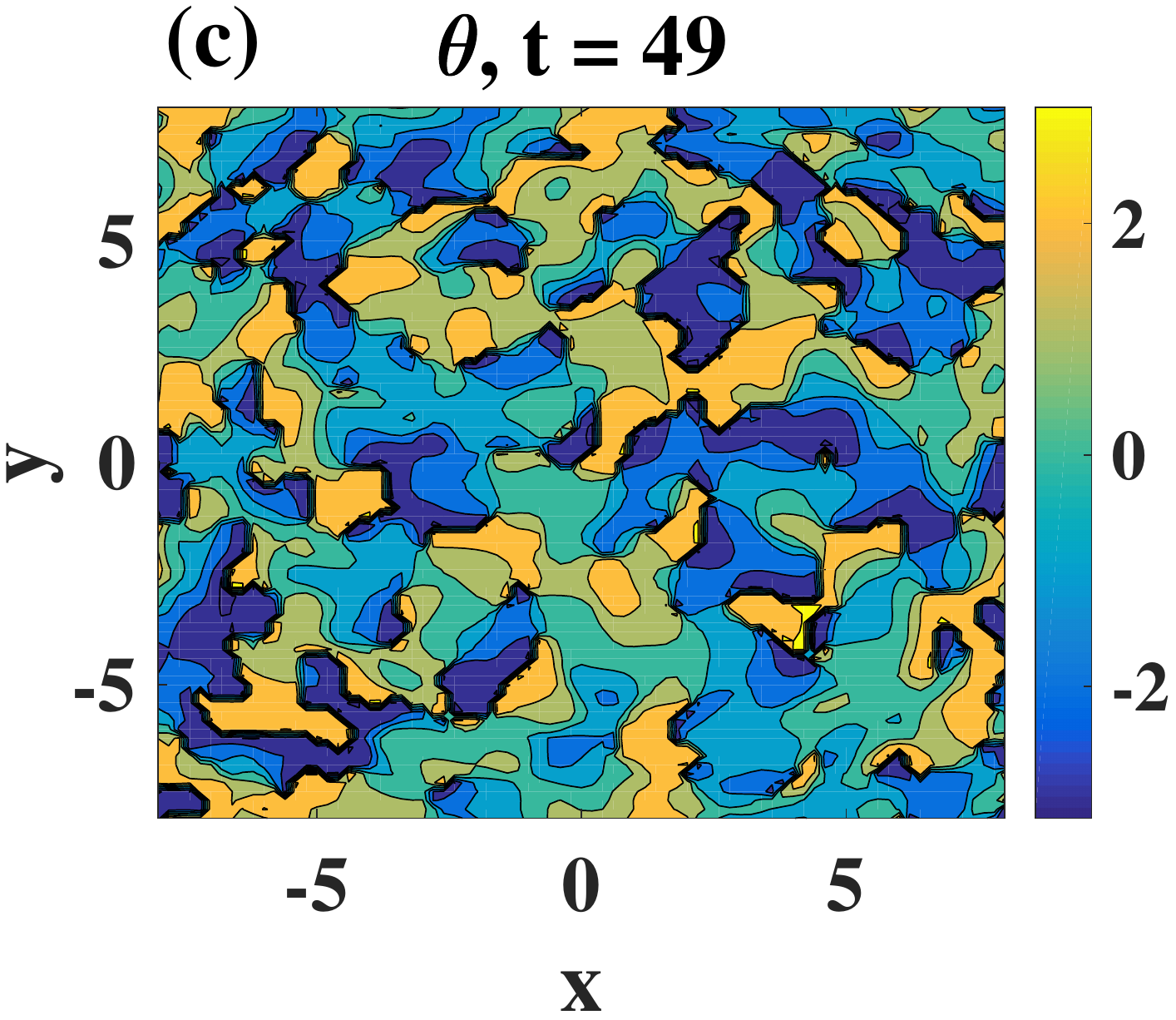}
	\includegraphics[width=0.22\textwidth]{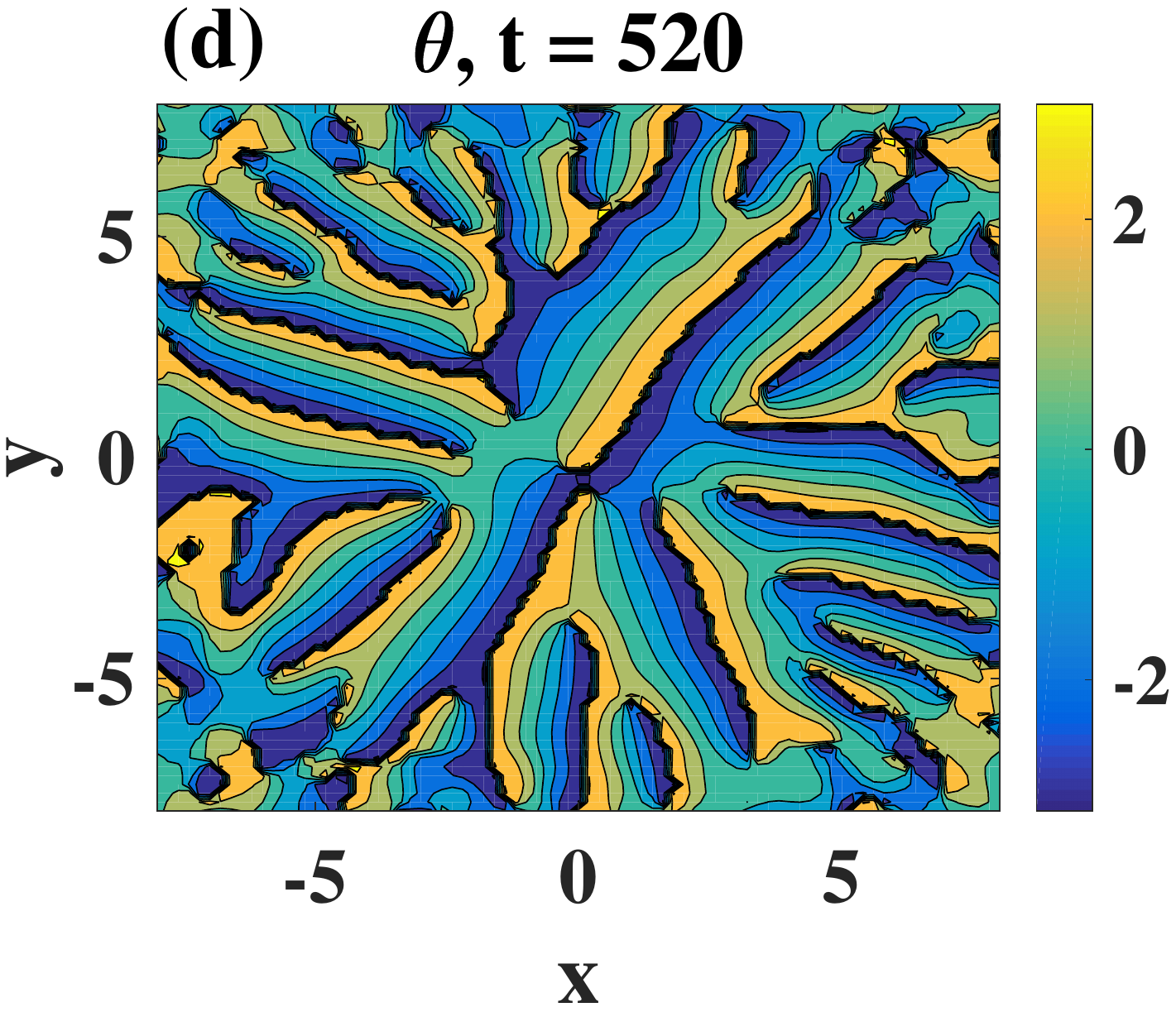}
	\caption{(Color online) Time evolution of polariton condensate described by ODGPE~(\ref{eqn:ODGPE_rescaled}) in the limit of $\gamma'_R\ll\gamma'_c$ with small interactions. Parameters are $g'_c=0.1$, $g'_R=0.1g'_c$, $\gamma'_c=1.26$, $\gamma'_R=0.1\gamma'_c$, and $R'=0.012\gamma'_c$. These figures depict the distributions of number density and phase of the polartiton condensate during time evolution as in Fig.~\ref{fig:case15}, where the laser radius and pumping power are chosen respectively as $R_P=5$ and $P'=8P'_{th}$. As the decay of exciton reservoir is very significant, the vortices configuration varies with time.}
	\label{fig:case19c}
\end{figure}

If we reduce the interactions so that they are negligible,~\cite{ohadi2016nontrivial} polariton condensates can be restored with spontaneous generated vortices, as depicted in Fig.~\ref{fig:case19c}.
In the limit of $\gamma'_R\ll\gamma'_c$, the low energy excitation spectrum deviates significantly from the linear Bogoliugov form, and presents a finite energy gap for instead~\cite{xu2017spinor}. However, from the results for $\gamma'_R = 0.1 \gamma'_c$ shown in Fig.~\ref{fig:case19c}, we can still observe spontaneously generated quantized vortices, without apparent lattice structure. In fact, the geometric configuration of vortices evolve with time, as one would expect for a liquid phase. We won't plot the number of vortices with various pumping radius $R_P$ and pumping power $P'$ for this case, since the number of vortices always change over time. 

\begin{figure}[tbp]
	\centering
	\includegraphics[width=0.22\textwidth]{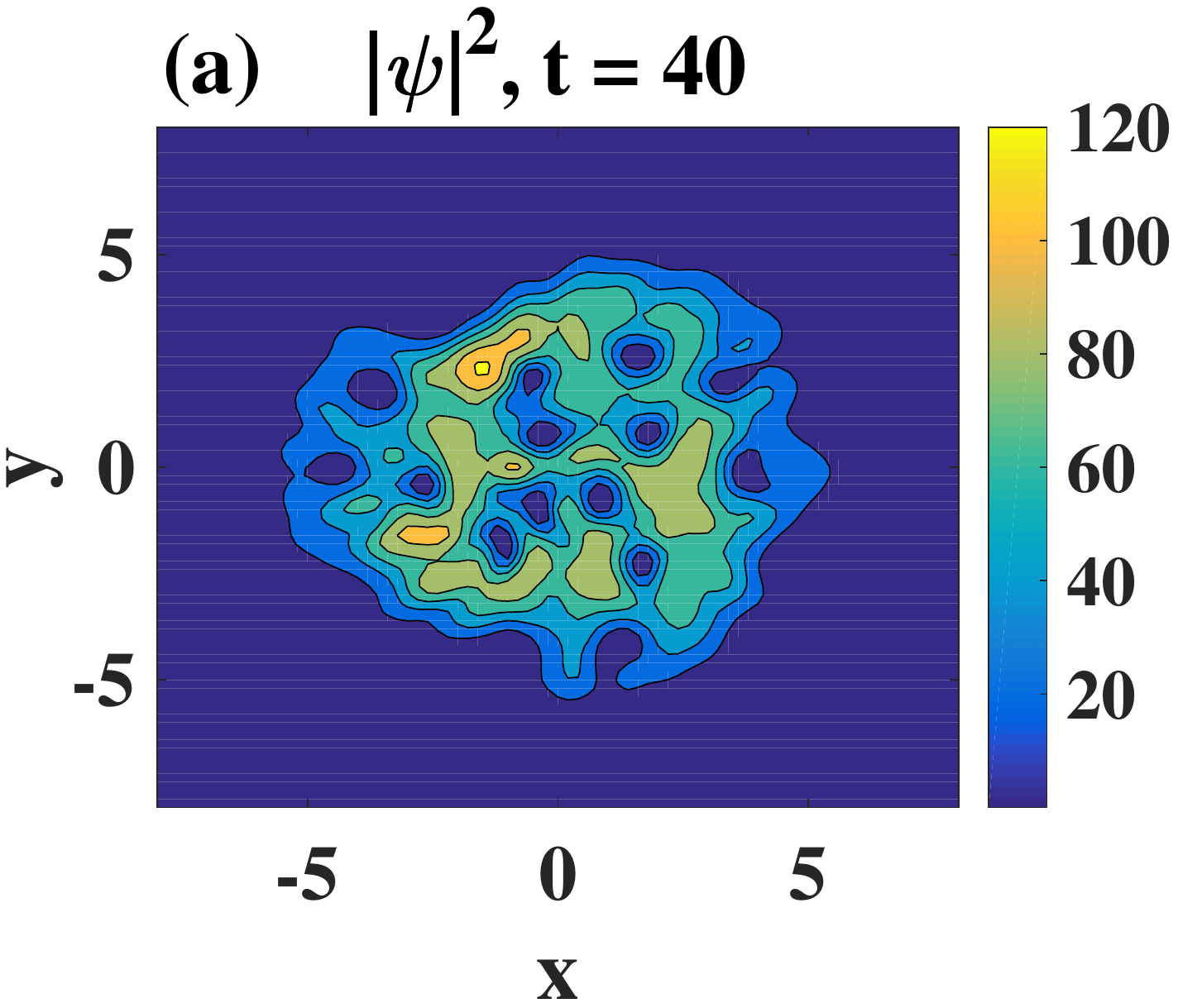}
	\includegraphics[width=0.22\textwidth]{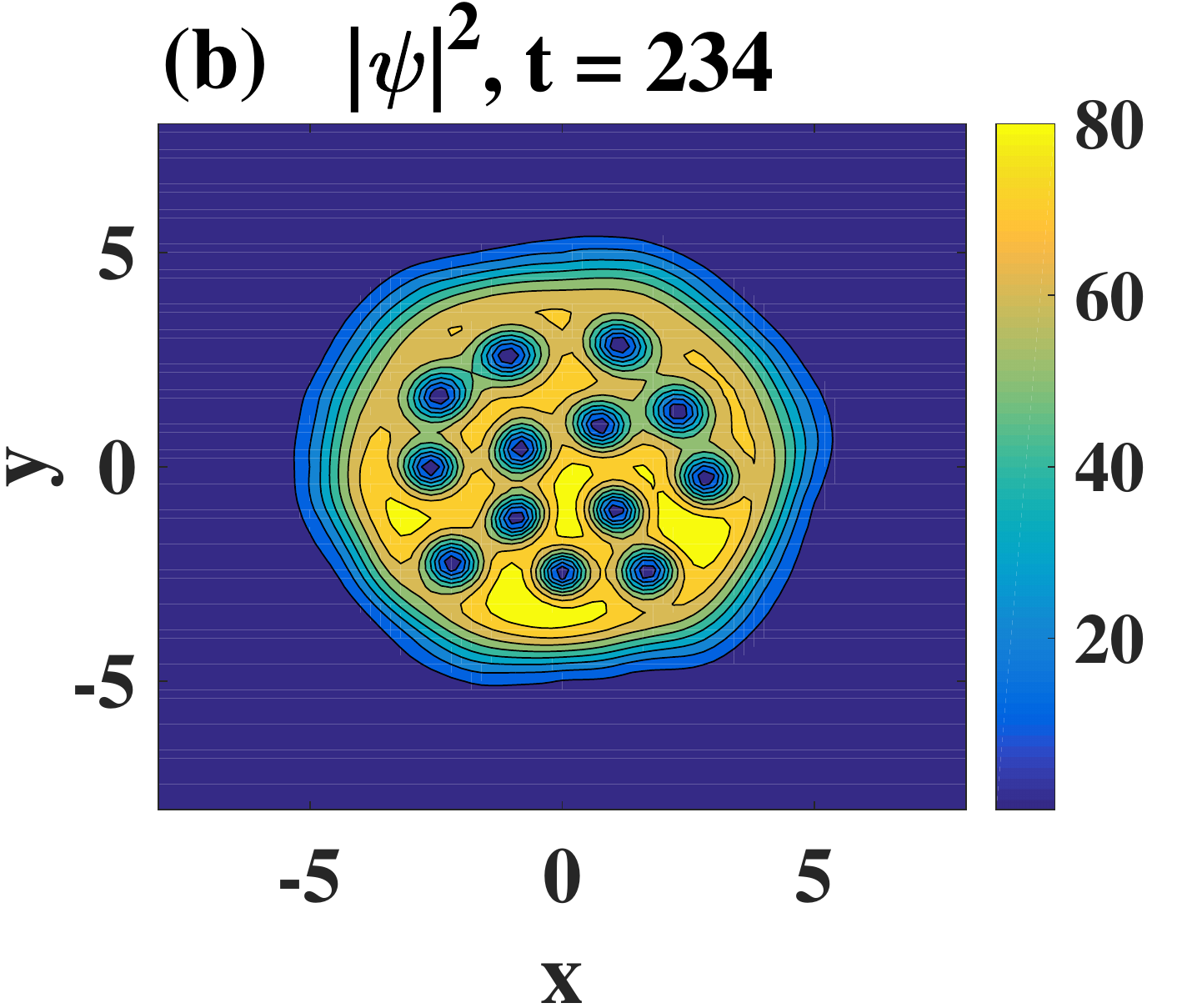}
	\\
	\vskip2mm
	\includegraphics[width=0.22\textwidth]{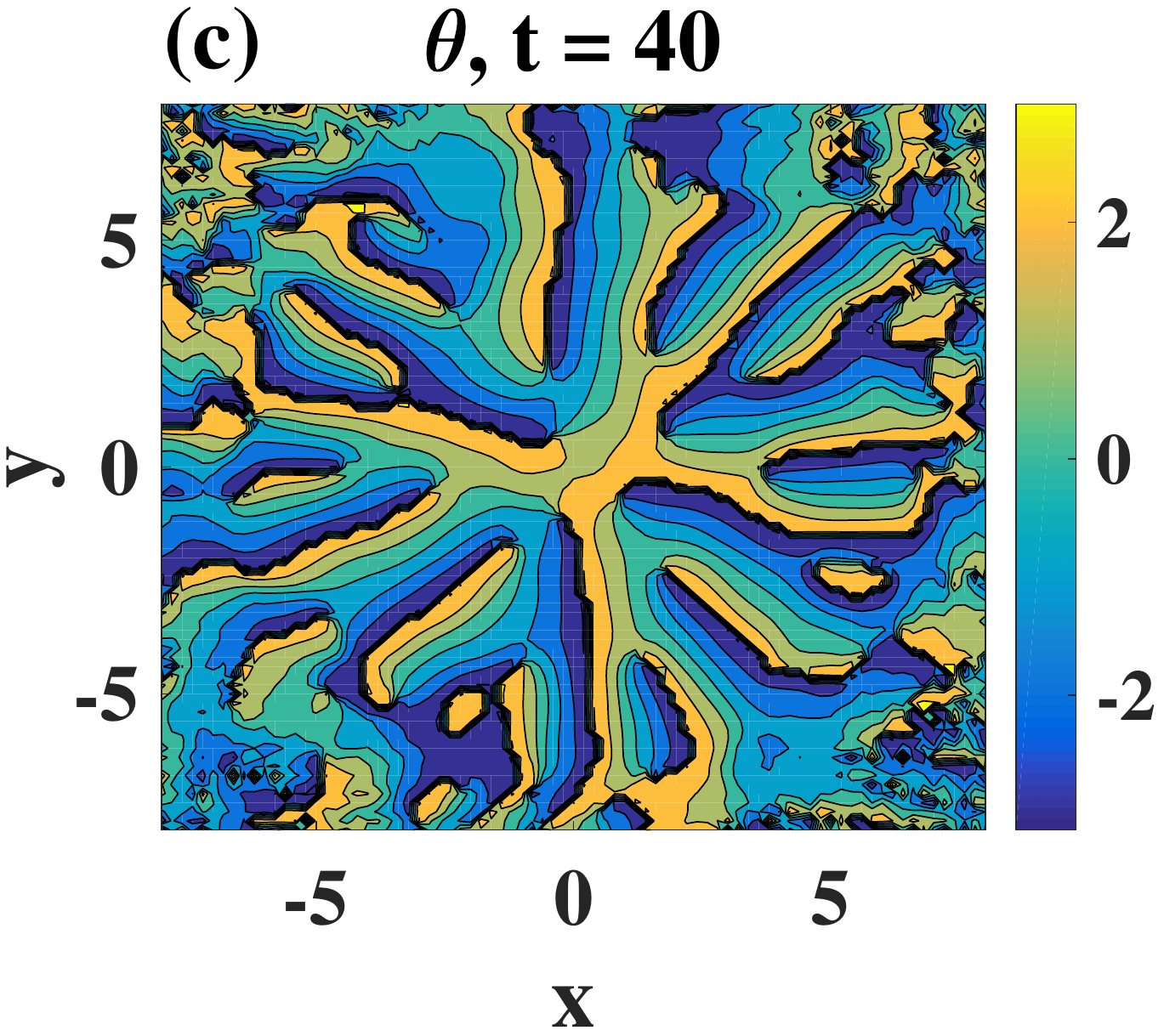}
	\includegraphics[width=0.22\textwidth]{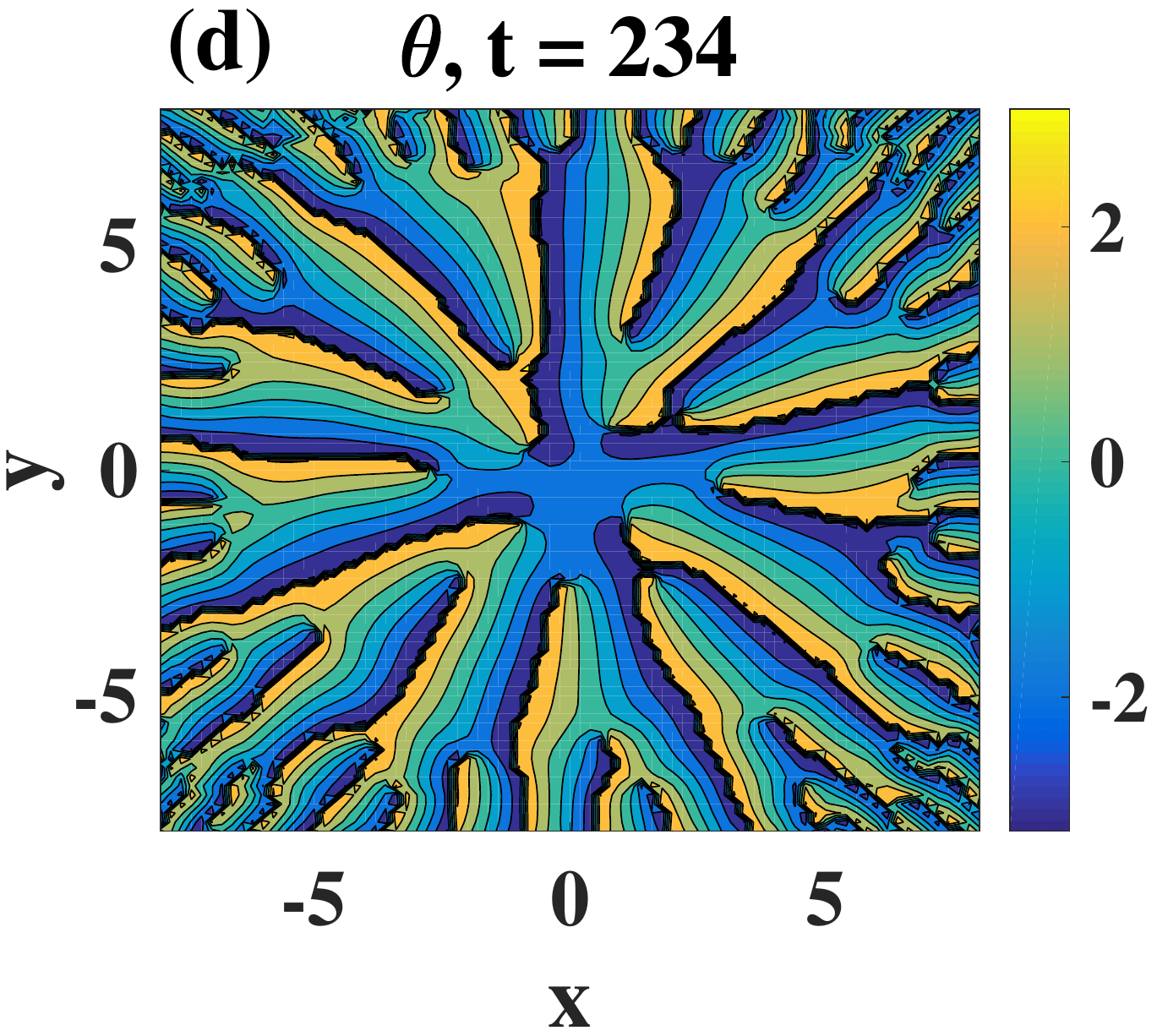}
	\\
	\vskip2mm
	\includegraphics[width=0.22\textwidth]{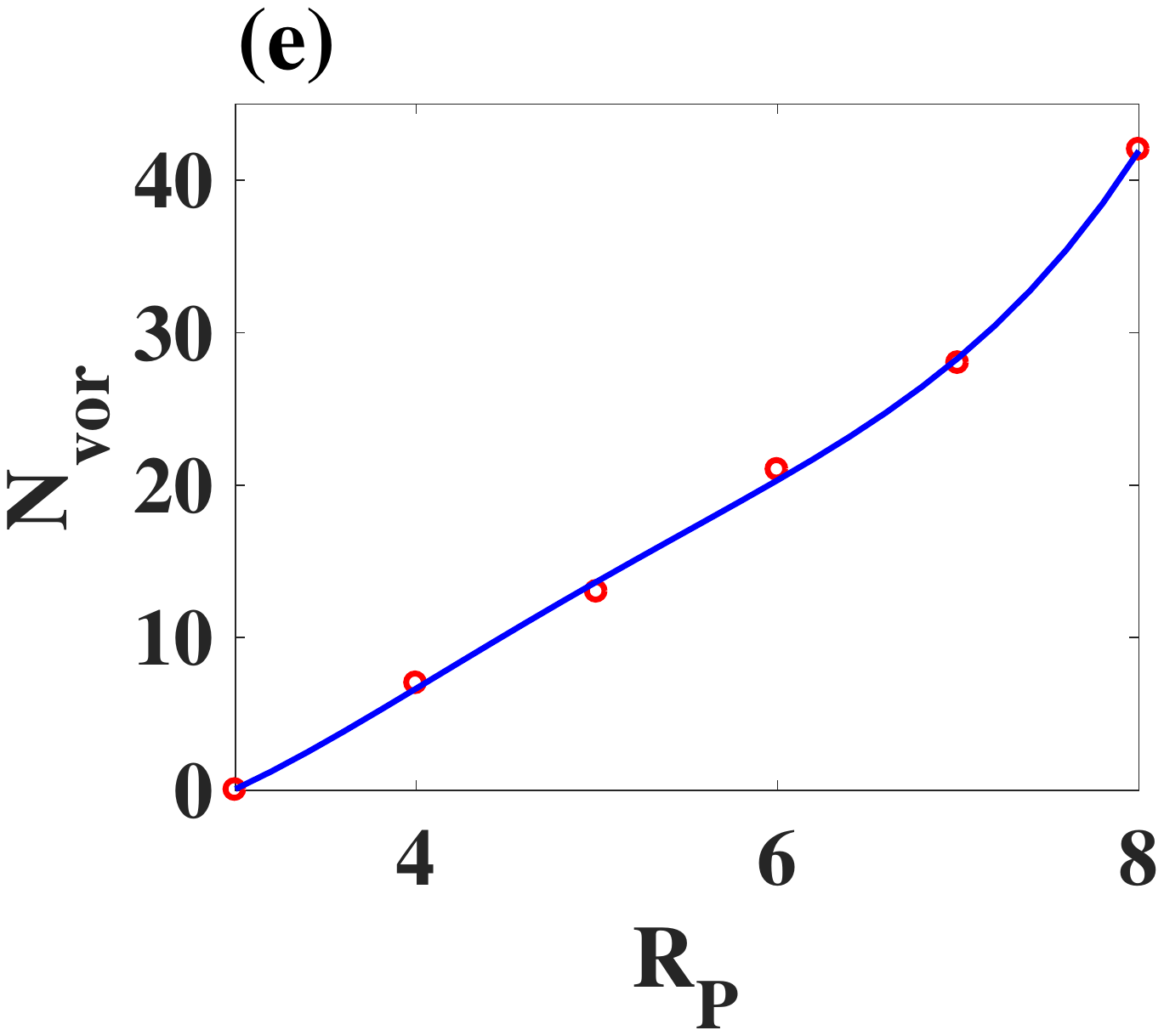}
	\includegraphics[width=0.22\textwidth]{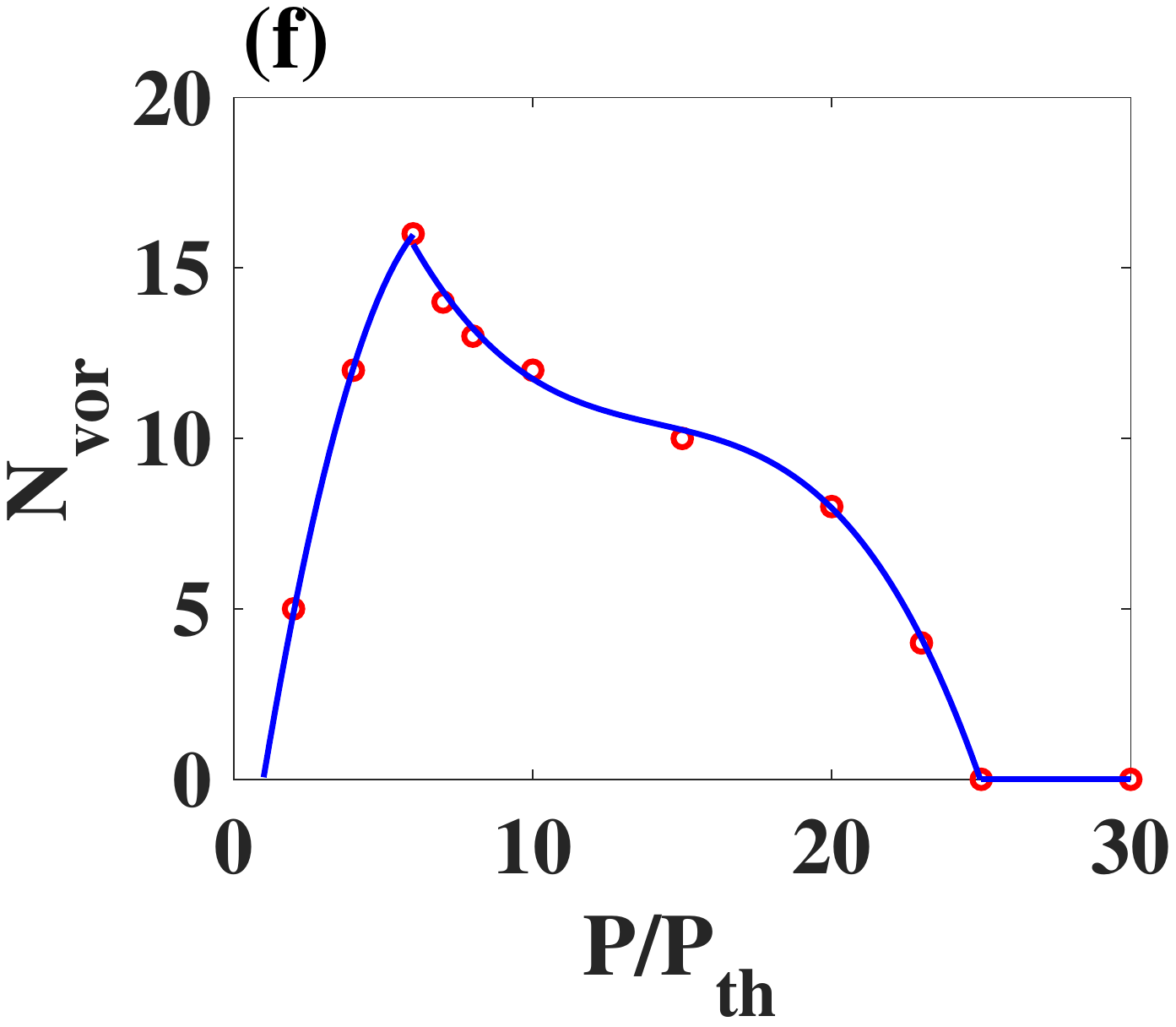}
	\caption{(Color online) Time evolution and steady-state solution of polariton condensates described by ODGPE~(\ref{eqn:ODGPE_rescaled}) in the limit of $\gamma'_R\ll\gamma'_c$ and subjected to an external unit angular momentum $\Omega=1$. Panels (a)-(d) display the same quantities as in Fig.~\ref{fig:case15} and parameters are chosen as in Fig.~\ref{fig:case19c}. The number of vortices in the steady states is depicted by changing (d) the pumping radius (f) the pumping power.}
	\label{fig:case19r}
\end{figure}

The numerical results discussed above suggest that quantized vortices can be generated in circularly pumped polariton BEC even for the case of $\gamma'_R\ll\gamma'_c$. However, the vortices are in a liquid phase with time-varying geometric configuration. This may hinder the direct observation of these vortices in experiments, as the polariton BEC is two-dimensional in nature, and the system is in the Berezinskii-Kosterlitz-Thouless (BKT) phase with spontaneously generated vortex and anti-vortex pairs.~\cite{roumpos2011single} It is then desirable to find a mechanism which can help stabilizing the lattice structure of vortices. Here, we apply an external rotation to the system by adding an angular momentum term $2\Omega L_z\psi$ into the first equation of the ODGPE [Eq.~(\ref{eqn:ODGPE_rescaled})], where $L_z=-i(x\partial_y-y\partial_x)$.~\cite{padhi2015vortex, chen2017quantum} 

In Fig.~\ref{fig:case19r}, we show results of a unit angular momentum $\Omega=1$ for the limit of $\gamma'_R = 0.1 \gamma'_c$, where the vortex lattice is completely melted without a rotation as shown in Fig.~\ref{fig:case19c}. 
By comparing these two cases, we find that although the number of vortices are only slightly enhanced by the external rotation, the lattice structure is perfectly restored as shown in Figs.~\ref{fig:case19r}(b) and \ref{fig:case19r}(d). This observation can be understood by noticing that an applied external rotation can effectively reinforce the boundary condition because the symmetry of single-particle wave function is consistent with the trapping potential. For a mesoscopic sample as considered here, the effect of boundary condition can be very influential in determining the configuration vortices. An external rotation can be implemented in experiment by either rotating the sample mechanically, or applying a Laguerre-Gauss laser beam with finite angular momentum.~\cite{phillips-06}

\section{Conclusion}
\label{sec:summary}

In this paper, we investigate the formation and stability of vortices and vortex lattices in exciton-polariton condensates under a non-coherent pumping laser of a circular cross section. Within a mean-field approach which takes into account the finite decay rates of exciton reservoir and polariton condensate, the time evolution of the BEC is obtained numerically to reveal the emergence and configuration of vortices in the long-time limit. By varying the decay rates of reservoir $\gamma'_R$ and condensate $\gamma'_c$, we systematically discuss different parameter regimes where adiabatic approximation is valid with $\gamma'_R\gg P'R'/2\gamma'_c$ and $\gamma'_R \gg \gamma'_c$, partly broken with $\gamma'_R  = P'R'/2\gamma'_c$ and $\gamma'_R \gg \gamma'_c$, fully broken with $\gamma'_R = \gamma'_c$, and broken to the opposite limit with $\gamma'_R \ll \gamma'_c$. We find that by gradually relaxing the adiabatic approximation in the sequence above, the number of vortices generated in the condensate is decreased, and the lattice structure starts to melt when $\gamma'_R \approx \gamma'_c$, and becomes completely liquified in the limit $\gamma'_R \ll \gamma'_c$. This observation suggests that the vortex lattice is more favorable and stable when the adiabatic approximation is valid. We further study the effect of an external rotation on the vortex generation and configuration, and find that an imposed angular momentum is very effective to stabilize the vortex lattice structure even in the limiting scenario $\gamma'_R \ll \gamma'_c$. Our results provide useful information to the study of superfluidity in condensates in quantum open systems. 

\begin{acknowledgements}
The authors acknowledge productive discussions with P. D. Drummond. This work is supported by the National Key R\&D Program of China (Grants No. 2016YFA0301302, No. 2018YFB1107200 and No. 2018YFA0306501), the National Natural Science Foundation of China (Grants No. 11522436, No. 11622428, No. 11434011, No. 11774425, No. 61475006, and No. 61675007), the Beijing Natural Science Foundation (Z180013), and the Research Funds of Renmin University of China (Grants No. 16XNLQ03 and No. 18XNLQ15). 

\end{acknowledgements}

\bibliography{VortexInPBEC}
\end{document}